\documentclass[%
 reprint,
 amsmath,amssymb,
 aps,
]{revtex4-1}

\usepackage{graphicx}
\usepackage{dcolumn}
\usepackage{bm}

\begin{document}

\preprint{APS/123-QED}

\title{ RE-ORIENTATION TRANSITION IN MOLECULAR THIN FILMS:
POTTS MODEL WITH DIPOLAR INTERACTION}

\author{Danh-Tai HOANG$^a$}%
 \email{danh-tai.hoang@u-cergy.fr}
\author{ Maciej KASPERSKI$^b$}%
 \email{maks@amu.edu.pl}
\author{Henryk PUSZKARSKI$^b$}%
 \email{henpusz@amu.edu.pl}
\author{H. T. DIEP$^a$}
\email{diep@u-cergy.fr, corresponding author}
\affiliation{%
$^a$Laboratoire de Physique Th\'eorique et Mod\'elisation,
Universit\'e de Cergy-Pontoise, CNRS, UMR 8089\\
2, Avenue Adolphe Chauvin, 95302 Cergy-Pontoise Cedex, France.\\
$^b$ Surface Physics Division, Faculty of Physics,
Adam Mickiewicz University\\
Umultowska 85, 61-614 Pozn\'an, Poland.\\}




\begin{abstract}
We study the low-temperature behavior and the phase transition of a thin film by Monte Carlo simulation. The thin film has a simple cubic lattice structure where each site is occupied by a Potts parameter which indicates the molecular orientation of the site. We take only three molecular orientations in this paper which correspond to the 3-state Potts model.  The Hamiltonian of the system includes: (i) the exchange interaction $J_{ij}$ between  nearest-neighbor sites $i$ and $j$  (ii) the long-range dipolar interaction of amplitude $D$ truncated at a cutoff distance $r_c$ (iii) a single-ion perpendicular anisotropy of amplitude $A$.
We allow $J_{ij} =J_s$ between surface spins, and $J_{ij}=J$ otherwise.  We show that the ground state depends on the  the ratio $D/A$ and $r_c$.  For a single layer, for a given $A$, there is a critical value $D_c$ below (above) which the ground-state (GS) configuration of molecular axes is perpendicular (parallel) to the film surface.  When the temperature $T$ is increased, a re-orientation transition occurs near $D_c$: the  low-$T$ in-plane ordering undergoes a transition to the perpendicular ordering at a finite $T$, below the transition to the paramagnetic phase.  The same phenomenon is observed in the case of a film with a thickness. We show that the surface phase transition can occur below or above the bulk transition depending on the ratio $J_s/J$.   Surface and bulk order parameters as well as other physical quantities are shown and discussed.

\begin{description}
\item[PACS numbers:64.60.De, 75.10.-b, 75.40.Mg, 75.70.Rf ]
\end{description}
\end{abstract}

\pacs{Valid PACS appear here}
\maketitle


\section{Introduction}

Surface physics has been intensively developed during the last 30 years.  Among the main reasons for that rapid and successful development we can mention the interest in understanding the physics of low-dimensional systems and an immense potential of industrial applications of thin films \cite{Zangwill,Bland,Diehl}.  In particular, theoretically it has been shown that systems of continuous spins (XY and Heisenberg) in two dimensions (2D) with short-range interaction cannot have long-range order at finite temperature  \cite{Mermin}.  In the case of thin films, it has been shown that low-lying localized spin waves can be found at the film surface \cite{Pusz} and effects of these localized modes on the surface magnetization at finite temperature ($T$) and on the critical temperature have been investigated by the Green's function technique \cite{Diep79,Diep81}.
Experimentally, objects of
nanometric size such as ultrathin films and nanoparticles have also been intensively studied because of numerous and important applications in industry.
An example is the so-called giant magnetoresistance
used in data storage devices, magnetic sensors, etc.
\cite{Baibich,Grunberg,Barthelemy,Tsymbal}.  Recently, much interest has been attracted towards practical problems such as spin transport, spin valves and spin-torques transfer, due to numerous applications in spintronics.

In this paper, we are interested  in the phase transition of the Potts model \cite{Baxter} in thin films taking into account a dipolar interaction and a perpendicular anisotropy.  The $q$-state Potts model is very popular in statistical physics and much is known for models with short-range ferromagnetic interactions in 2D and 3D \cite{Baxter}. The Potts model with an algebraically decaying long-range interaction has been investigated in 1D \cite{Bayong,Reynal}. Such a monotonous long-range interaction can induce an ordering at finite $T$ in one-dimensional systems.   The dipolar interaction, however,  is very special because it contains two competing terms which yield complicated orderings depending on the sample shape.  For example, the dipolar interaction favors an in-plane ordering in films and slabs with infinite lateral dimensions.  Many studies have been done with the dipolar interaction in thin films with the Heisenberg spin model \cite{Pusz2008,Santamaria}.  The absence of the Potts model for thin films has motivated the present work.

We will consider a thin film made of a molecular crystal where molecular spins can point along the $x$, $y$ or $z$ axes.  The interactions between molecular spins include  a dipolar interaction truncated at a distance $r_c$ and an exchange interaction between nearest neighbors (NN). We  also take into account a single-ion perpendicular anisotropy which is known to exist in ultrathin films \cite{Zangwill}.  The method we employ is Monte Carlo (MC) simulations.  Phase transition in systems of interacting particles is a major domain in statistical physics. Much is now understood with the analysis provided by the fundamental concepts of the renormalization group \cite{Wilson} and with the use of the field theory \cite{Zinn}.  But these methods encountered some difficulties in dealing with frustrated spin systems \cite{Diep2005,Diep1991}.  MC simulations are therefore very useful to complete theories and to interpret experiments. They serve as testing means for new theoretical developments. Over the years, the standard MC method \cite{Binder} has been improved by the finite-size scaling theory \cite{Hohenberg} and by other  high-performance techniques  such as histogram techniques \cite{Ferrenberg1,Ferrenberg2}, cluster updating algorithms \cite{Hoshen,Wolff,Wolff2} and Wang-Landau flat-histogram method \cite{Wang-Landau}.  We have now at hand these efficient techniques to deal with complex systems. We can mention our recent investigations by MC techniques on multilayers \cite{Ngo2004}, on frustrated surfaces \cite{Ngo2007,Ngo2007a} or on surface criticality \cite{Pham1,Pham2}.

In section II, we describe our model and the method we employ.
  Results of MC simulations are shown  and discussed in section III for several cases: 2D, homogeneous films and effects of surface interaction.
Concluding remarks are given in section \ref{conclu}.

\section{Model and Method}\label{model}

We consider a thin film of simple cubic lattice.  The film is infinite in the $xy$ plane and has a thickness $L_z$ in the $z$ direction.  The  Hamiltonian is given by the following 3-state Potts model:

\begin{equation}\label{HL}
{\cal H} = -\sum_{(i,j)}J_{ij}\delta(\sigma_{i},\sigma_{j} )
\end{equation}
where $\sigma_{i}$ is a variable associated to the lattice site $i$.  $\sigma_{i}$ is equal to 1, 2 and 3 if the spin at that site lies along the $x$,  $y$ and $z$ axes, respectively. $J_{ij}$ is the exchange interaction between NN at $i$ and $j$.  We will assume that (i) $J_{ij}=J_s$ if $i$ and $j$ are on the same film surface (ii) $J_{ij}=J$ otherwise.

The dipolar Hamiltonian is written as

\begin{eqnarray}
{\cal H}_d&=&D\sum_{(i,j)}\{\frac{\mathbf{S}(\sigma_{i})\cdot \mathbf{S}(\sigma_{j})}{r_{i,j}^3}\nonumber \\
&&-3\frac{[\mathbf{S}(\sigma_{i})\cdot \mathbf r_{i,j}][\mathbf{S}(\sigma_{j})\cdot \mathbf r_{i,j}]}{r_{i,j}^5}\}
\label{dip}
\end{eqnarray}
where $\mathbf r_{i,j}$ is the vector  of modulus $r_{i,j}$  connecting the site  $i$ to the site $j$. One has  $\mathbf r_{i,j}\equiv \mathbf r_j-\mathbf r_i$.   In Eq. (\ref{dip}), $D$ is a positive constant depending on the material, the sum $\sum_{(i,j)}$
is limited at pairs of spins within a cut-off distance $r_c$, and $\mathbf{S}(\sigma_{i})$ is given by the following three-component pseudo vector representing the spin state
\begin{eqnarray}
\mathbf{S}(\sigma_{i})&=&(s_x(i),0,0) \  \  \mbox{if}  \  \   \sigma_{i}=1\\
\mathbf{S}(\sigma_{i})&=&(0,s_y(i),0)  \  \  \mbox{if}  \  \   \sigma_{i}=2\\
\mathbf{S}(\sigma_{i})&=&(0,0,s_z(i))  \  \    \mbox{if}  \  \   \sigma_{i}=3
\end{eqnarray}
where $s_\alpha$ ($\alpha=x,y,z$) is the $\alpha$ component with values $\pm 1$.

The perpendicular anisotropy is introduced by the following term
\begin{equation}\label{HA}
{\cal H}_a = -A\sum_{i}s_{z}(i)^2
\end{equation}
where $A$ is a constant.

Note that the dipolar interaction as applied in our Potts model is not similar to that used in the vector spin model where $\mathbf S(\sigma_{i})$ is a true vector.  In our model, each spin can only choose to lie on one of three axes, pointing in positive or negative direction.

We use $J=1$ as the unit of energy.  The temperature $T$ is expressed in the unit of $J/k_B$ where $k_B$ is the Boltzmann constant.

In the absence of $D$, the GS configuration is perpendicular to the film surface due to the term ${\cal H}_a$.  In the absence of $A$, the GS is an in-plane configuration due to $D$.  When both $A$ and $D$ are present, the GS depends on the ratio $D/A$.   An analytical determination of the GS is impossible due to the long-range interaction. We therefore determine the GS by the numerical steepest-descent method which works very well in systems with uniformly distributed interactions.  This method is very simple\cite{Ngo2007,Ngo2007a} (i) we generate a random initial spin configuration  (ii) we calculate the local field created at a given spin by its neighbors using Eqs. (\ref{HL}) and   (\ref{dip}) (iii) we change the spin axis to minimize its energy (i. e. we align the spin in its local field) (iv) we go to another spin and repeat until all spins are visited: we say we make one sweep (v) we  do a large number of sweeps until a good convergence to the lowest energy is reached.

We shall use MC simulation to calculate properties of the system at finite  $T$.  Periodic boundary conditions are used in the $xy$ planes for sample sizes of $L\times L\times L_z$ where $L_z$ is the film thickness.  Free symmetric surfaces are supposed for simplicity. Standard MC method \cite{Binder} is used to get general features of the phase transition. Systematic finite-size scaling to obtain critical exponents is not the purpose of the present work. In general, we discard several millions of MC steps per spin to equilibrate the system before averaging physical quantities over several millions of MC steps.
The averaged energy and the specific heat are defined by

\begin{eqnarray}
 \langle U\rangle&=&<{\cal H} + {\cal H}_d+{\cal H}_a>\\
C_V&=&\frac{\langle U^2\rangle-\langle U\rangle^2}{k_BT^2}
\end{eqnarray}
where $<...>$ indicates the thermal average taken over several millions of microscopic states at $T$.

We define the order parameter $Q$  for the $q$-state Potts model by
\begin{equation}\label{Q}
Q=[q\max (Q_1,Q_2,Q_3)-1]/(q-1)
\end{equation}
where $Q_n$ is the spatial average defined by
\begin{equation}\label{Qn}
Q_n=\sum_j \delta (\sigma_{i}-n)/(L\times  L\times L_z)
\end{equation}
$n(n=1,2,3)$ being the value attributed to denote the axis of the spin $\sigma_{i}$ at the site  $i$.
The susceptibility is defined by
\begin{equation}\label{chi}
\chi=\frac{\langle Q^2\rangle-\langle Q\rangle^2}{k_BT}
\end{equation}

We did not use the theory of finite-size scaling\cite{Hohenberg,Ferrenberg1,Ferrenberg2} because the calculation of critical exponents is not the purpose of the present work. However, in order to appreciate finite-size effects,  we carried out simulations in the 2D case for sizes from  $L\times L=24\times24$ to $60\times60$ and in the case of thin films from $L\times L\times L_z=12\times12\times 4$ to $48\times 48\times 6$.  Results for the largest size are not significatively different from those of smaller sizes, excepted for the thickness. We will show therefore in the following results for lateral lattice size $L=60$ for the 2D case, and $L=24$ for thin films with thicknesses $L_z=4$ and 6.
In order to check the first-order nature of a weak first-order transition, the histogram technique is very efficient \cite{Ferrenberg1,Ferrenberg2}. But in our case as will be seen below, the re-orientation  is a very strong first-order transition. The discontinuity of energy and magnetization is clearly seen at the transition. We just use the histogram technique to check the 2D case for a demonstration.

\section{Ground State and Phase Transition}
\subsection{Two dimensions}

\begin{figure}
\centering
\includegraphics[width=80mm,angle=0]{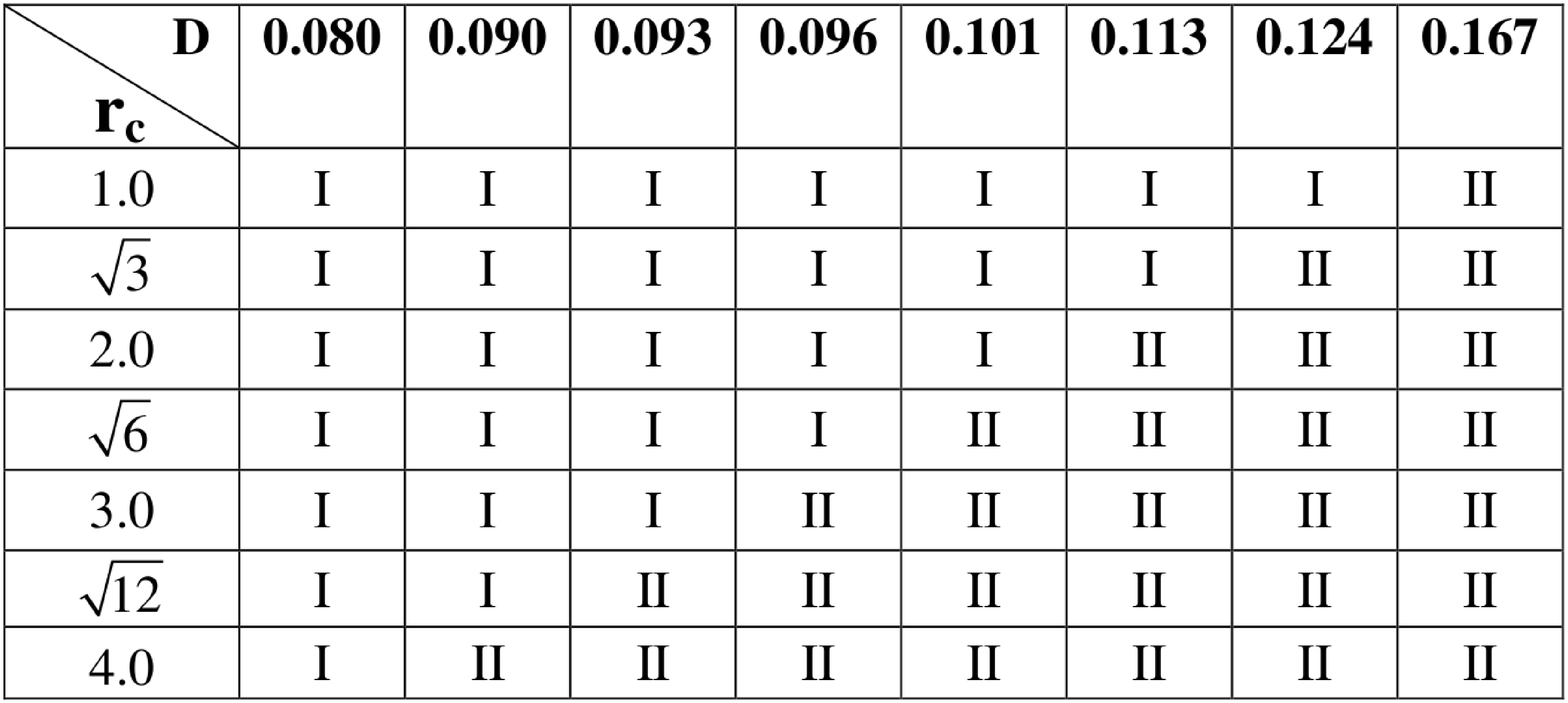}
\caption{Ground states as functions of $(D,r_c)$, with $A=0.5$, $J=1$: the number (I) stands for the perpendicular configuration and the number (II) for the in-plane configuration (spins pointing along $x$ or $y$ axis).} \label{GS2D}
\end{figure}

In the case of 2D, for a given $A$, the steepest-descent method gives the "critical value" $D_c$  of $D$ above (below) which the GS is the in-plane (perpendicular) configuration.  $D_c$ depends on  $r_c$.  Let us take $A=0.5$ and make vary $D$ and $r_c$ in the following.  The GS numerically obtained is shown in Fig. \ref{GS2D} for several sets of $(D,r_c)$.  For instance, when $r_c=\sqrt{6}\simeq 2.449$, we have $D_c=0.100$.

\begin{figure}
\centering
\includegraphics[width=50mm,angle=0]{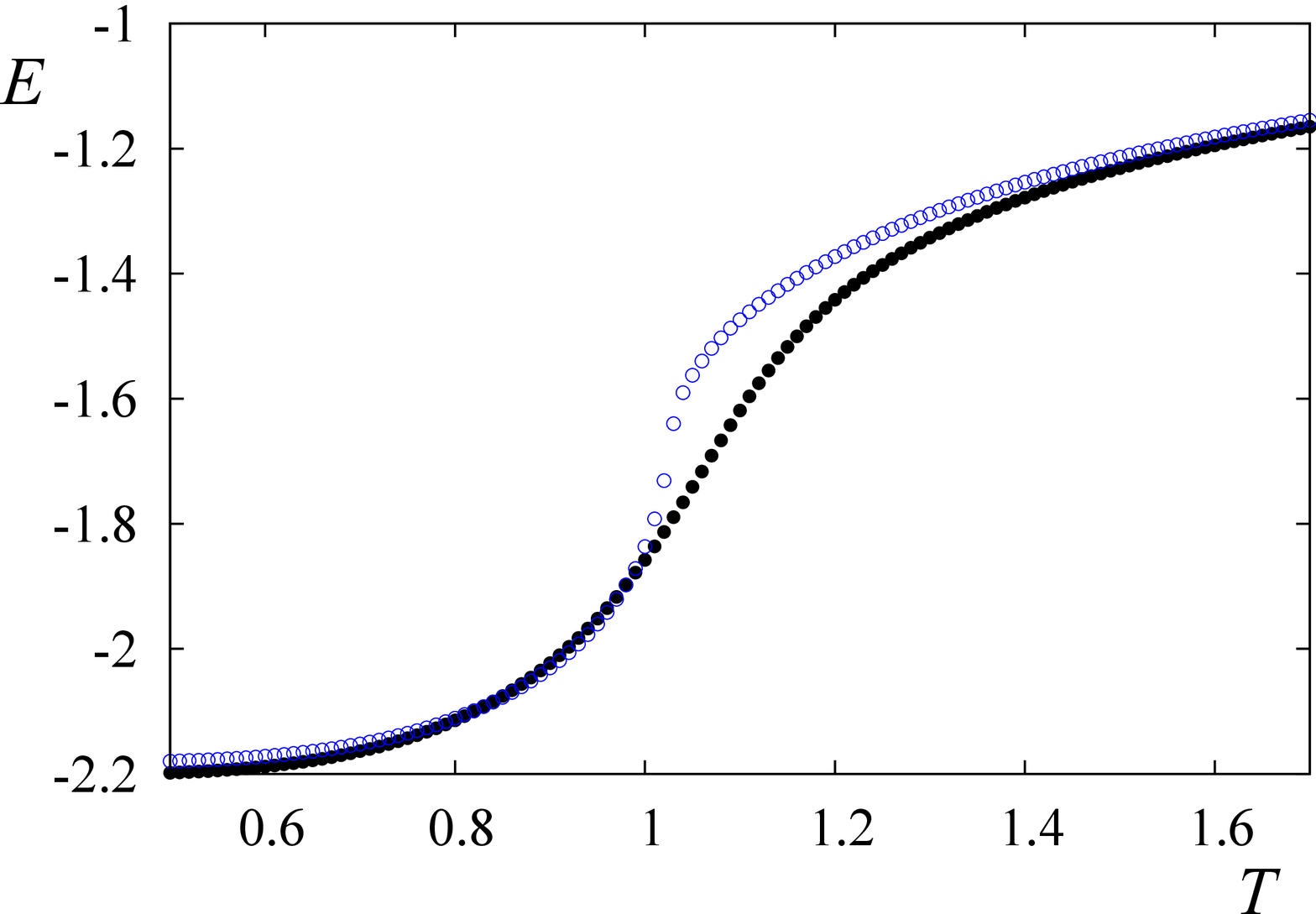}
\includegraphics[width=50mm,angle=0]{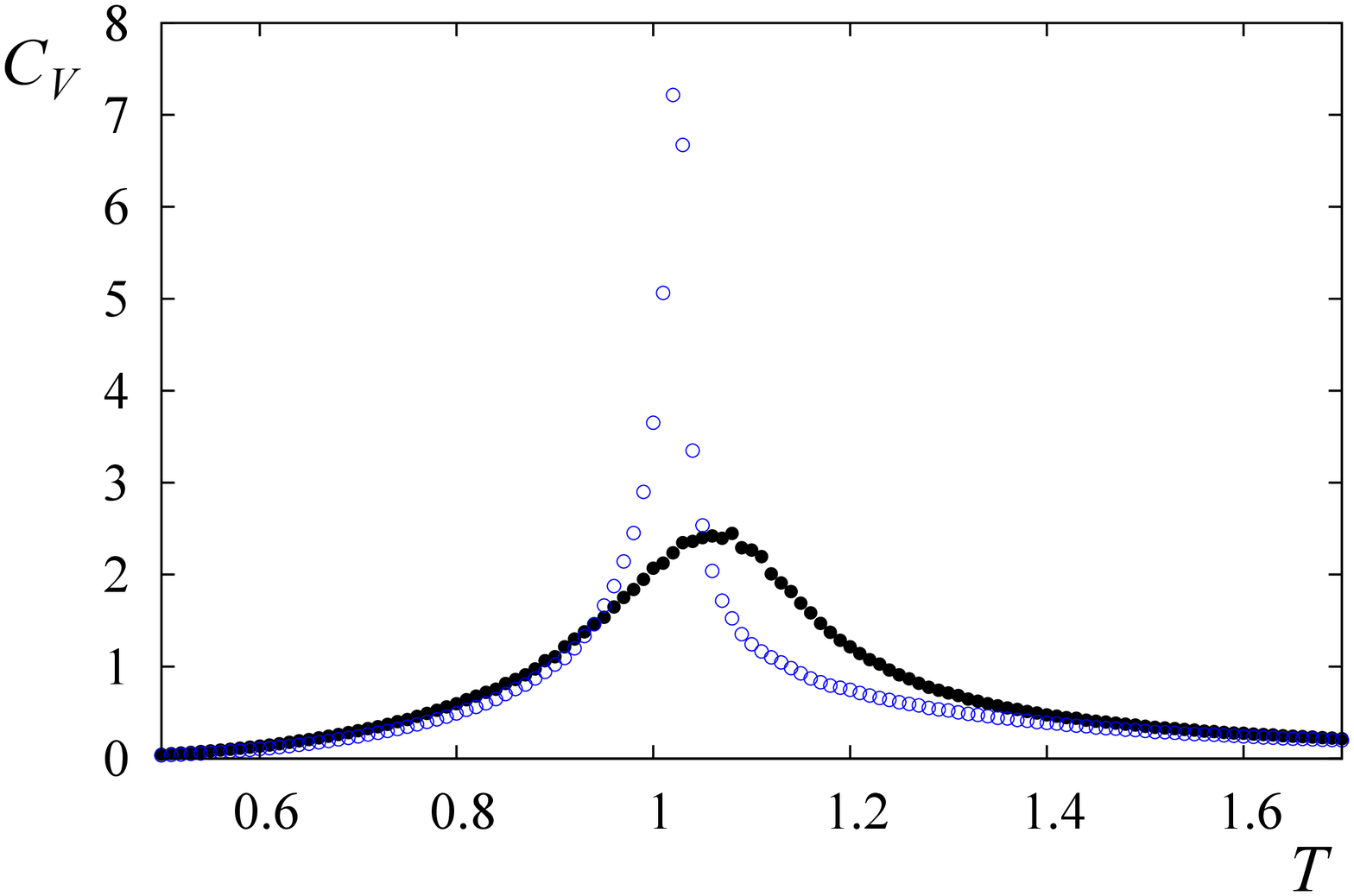}
\caption{(Color online) Energy $E$ and specific heat $C_V$ versus $T$ for $D=$0.09 (black solid circles) and 0.11(blue void circles), $L=60$, $A=0.5$, $J=1$, $r_c=\sqrt{6}$. } \label{EC}
\end{figure}

\begin{figure}
\centering
\includegraphics[width=50mm,angle=0]{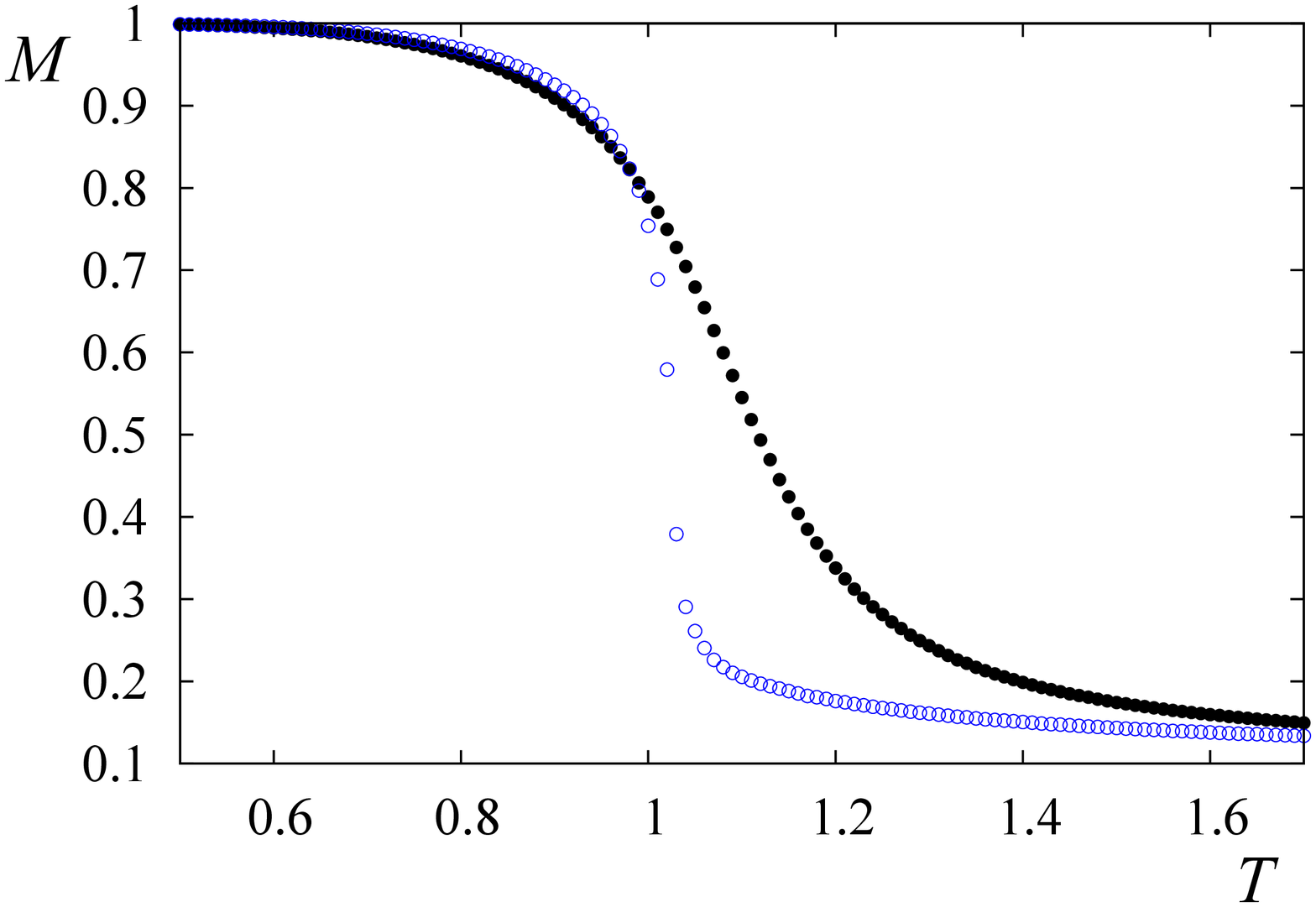}
\includegraphics[width=50mm,angle=0]{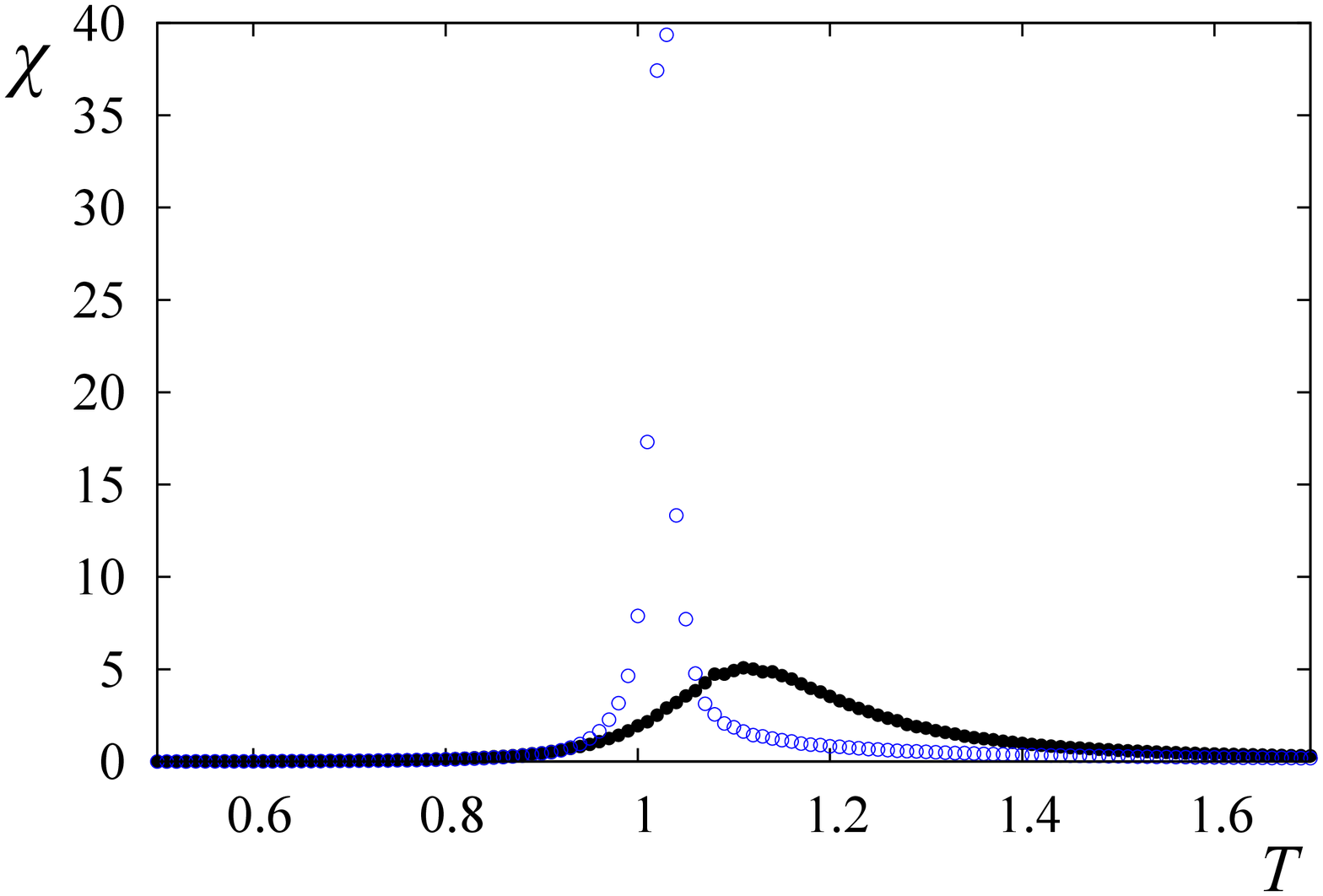}
\caption{(Color online) $M$ and $\chi$ versus $T$ for $D=$0.09 (black solid circles) and 0.11(blue void circles).  Note that $M$ for $D=0.09$ is the perpendicular magnetization while $M$ for $D=0.11$ is the in-plane magnetization, $L=60$, $A=0.5$, $J=1$, $r_c=\sqrt{6}$.} \label{MX}
\end{figure}

We show in Fig. \ref{EC}  the energy per site $E\equiv <U>/(L\times L\times L_z)$ and the specific heat, and in Fig. \ref{MX}
the order parameter $M=<Q>$ as well as the susceptibility $\chi$, as functions of $T$ in the case of $r_c=\sqrt{6}$,  for  $D=0.09$ and $D=0.11$ on two sides of $D_c=0.100$.  We observe one transition of second order for these values of $D$. Note that the transition for larger $D$ is sharper.

\begin{figure}
\centering
\includegraphics[width=50mm,angle=0]{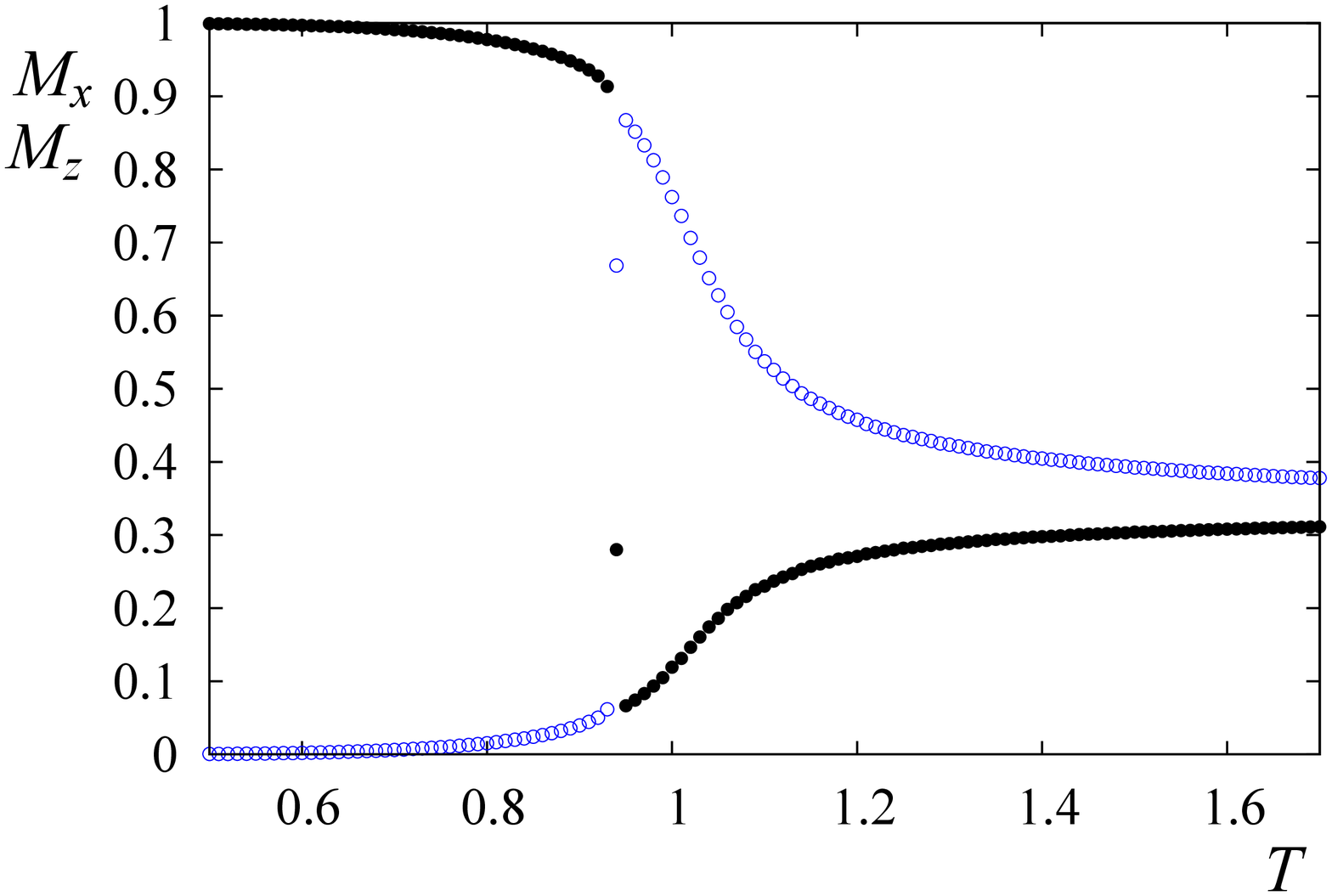}
\includegraphics[width=50mm,angle=0]{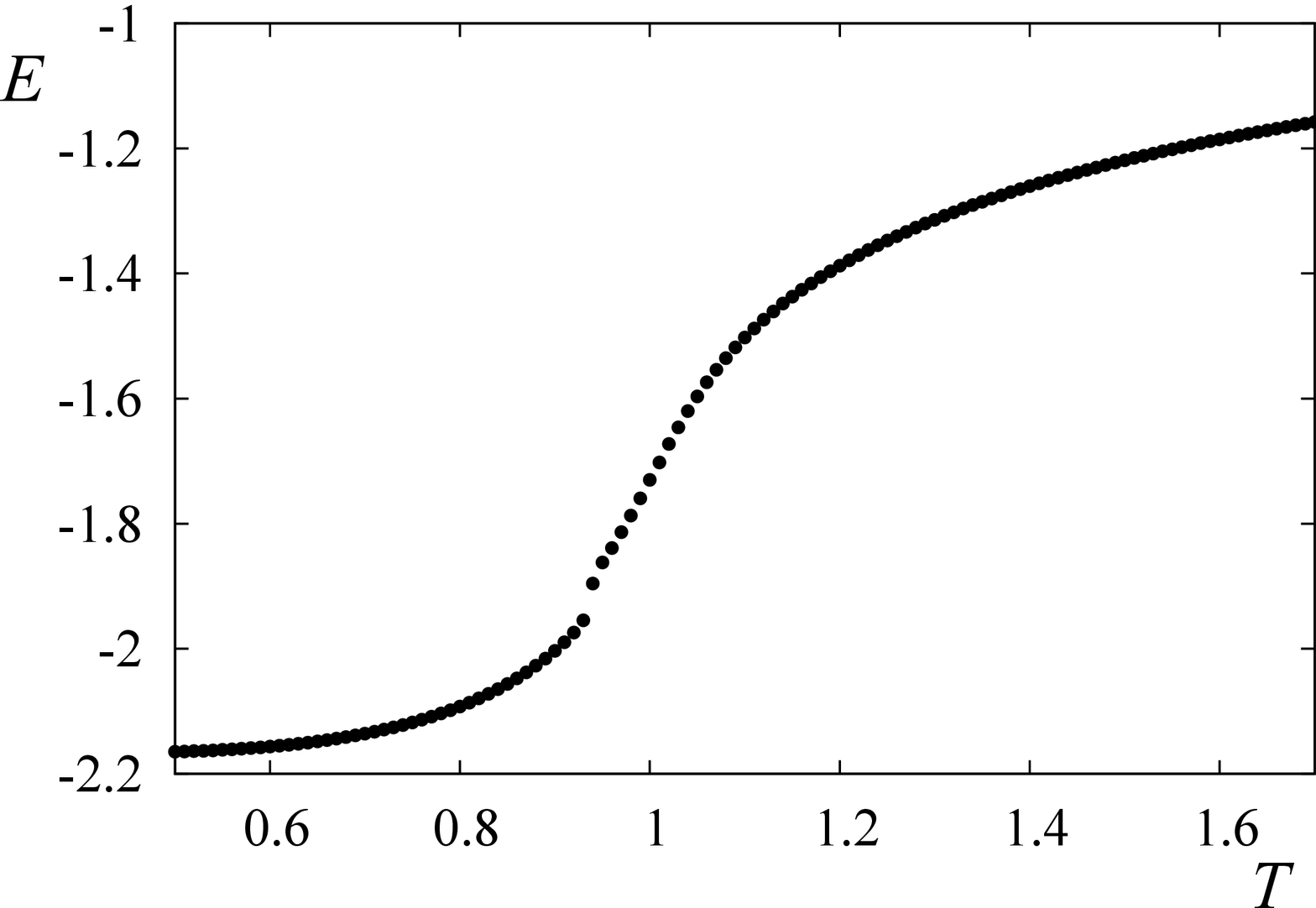}
\includegraphics[width=50mm,angle=0]{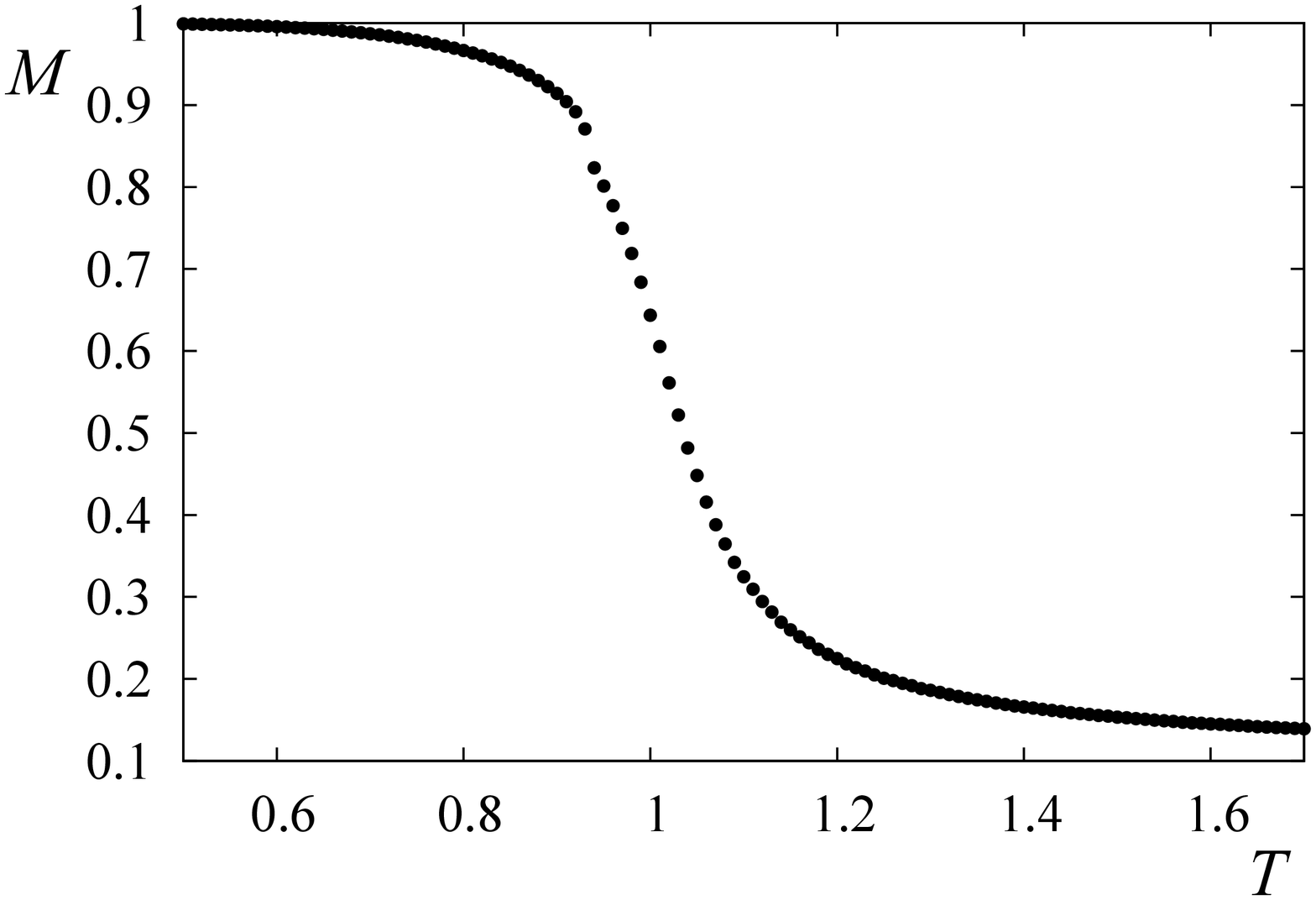}
\caption{(Color online) Energy per spin $E$, total magnetization $M$, $M_x$ (black solid circles) and $M_z$ (blue void circles) versus $T$ for $D=$0.101 in the re-orientation transition region, $L=60$, $A=0.5$, $J=1$, $r_c=\sqrt{6}$.} \label{MXMZ}
\end{figure}

\begin{figure}
\centering
\includegraphics[width=50mm,angle=0]{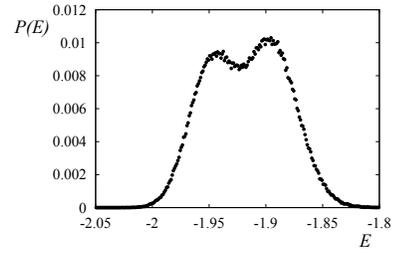}
\caption{ Energy histogram $P$ versus energy $E$ at the re-orientation transition temperature $T=0.930$, for $D=$0.101, $A=0.5$, $J=1$, $r_c=\sqrt{6}$ ($L=60$).} \label{2D-P}
\end{figure}

It is interesting to examine the region very close to $D_c$, namely close to the frontier of two different GS.  We have seen in the past that many interesting phenomena occur at the boundaries of different phases: we can mention the reentrance phenomenon in frustrated spin systems \cite{Diep2005,Diep1991} and the re-orientation transition in the Heisenberg film with a dipolar interaction similar to the present model \cite{Santamaria}.
We have carried out simulation for values close to $D_c$.  We find indeed a transition from the in-plane ordering to the perpendicular one when $T$ increases in the region $D \in [0.100,0.104]$. We show an example at $D=0.101$ in Fig. \ref{MXMZ} where we observe that in the low-$T$ phase ($0\leq T<0.93$) the spins align parallel to the $x$ axis and in the intermediate-$T$ phase ($0.93<T<1.05$) the spins point along the $z$ axis perpendicular to the film. The system becomes disordered at $T>1.05$.  Note that in the disordered phase, each "state" of the Potts spin (along of one of the three axes) has 1/3 of the total number of spins. This explains why $M_x$ and $M_z$ tend to 1/3 at high $T$ in Fig. \ref{MXMZ}. The transition from the in-plane to the perpendicular configuration is of first order as seen in Fig. \ref{MXMZ} by the discontinuity of $M_x$, $M_z$, the energy and the magnetization at the transition point.  The first-order character has been confirmed by the double-peaked energy histogram at the re-orientation transition temperature as shown in Fig. \ref{2D-P}.

\begin{figure}
\centering
\includegraphics[width=70mm,angle=0]{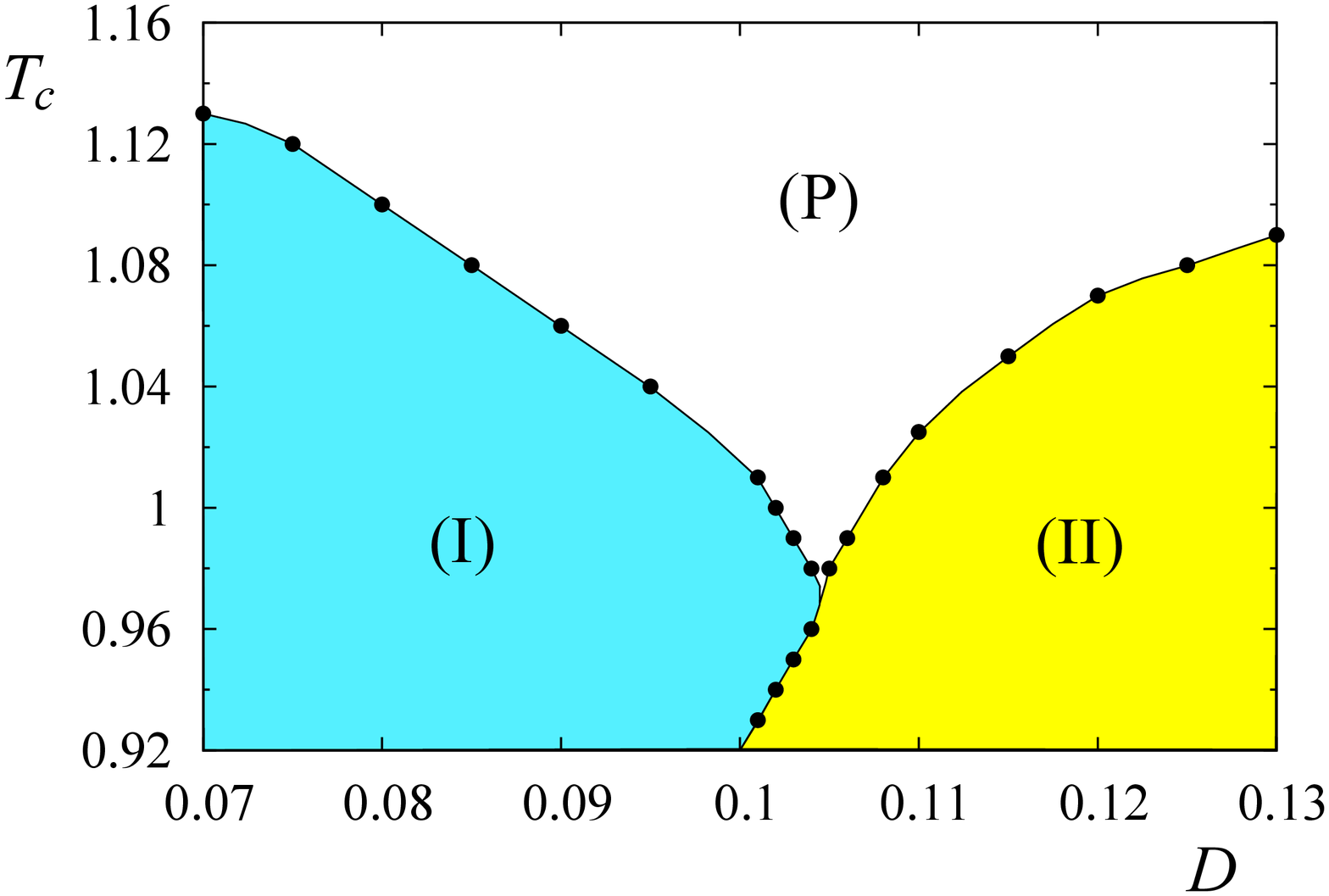}
\includegraphics[width=70mm,angle=0]{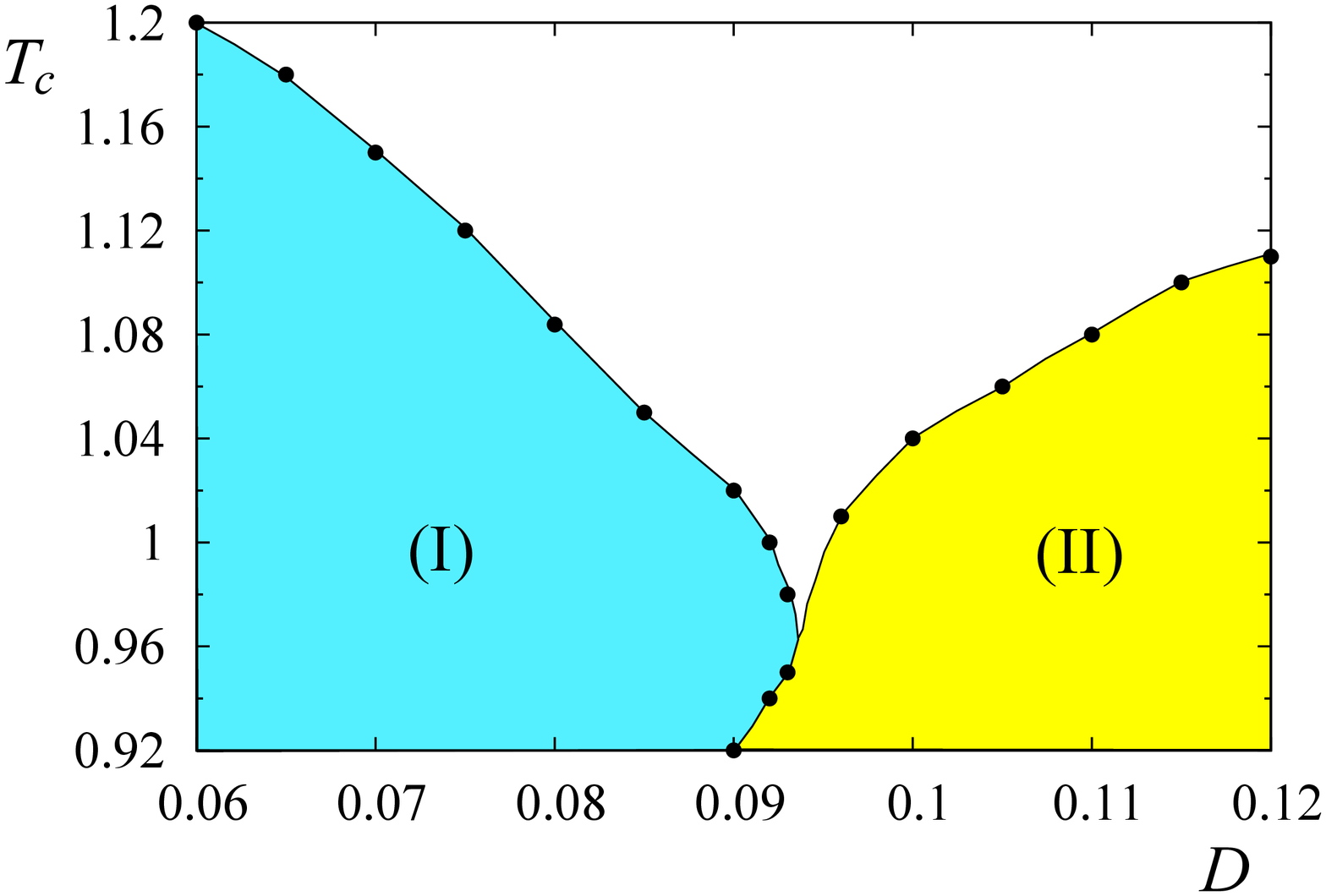}
\caption{(Color online) Phase diagram in 2D: Transition temperature $T_C$ versus $D$, with $A=0.5$, $J=1$, $r_c=\sqrt{6}$ (top) and $r_c=4$ (bottom). Phase (I) is the perpendicular spin configuration, phase (II) the in-plane spin configuration and phase (P) the paramagnetic phase. See text for comments. } \label{PD2D}
\end{figure}

We show in Fig. \ref{PD2D} (top) the phase diagram in the space $(D,T)$ for $r_c=\sqrt{6}$ where the line of re-orientation transition near $D_c$ is a line of first order.  Let us discuss about the effect of changing $r_c$. Increasing $r_c$ will increase the dipolar energy at each site. Therefore, a smaller value of $D$ suffices to "neutralize" the effect of perpendicular anisotropy energy. The critical value of $D_c$ is thus reduced as seen in the phase diagram established with $r_c=4$ shown in Fig. \ref{PD2D} (bottom) where $D_c=0.090$ compared to $D_c=0.100$ when $r_c=\sqrt{6}$ (top).

It is interesting to compare the present system using the 3-state Potts model with the same system using the Heisenberg spins \cite{Santamaria}.  In that work, the re-orientation transition line is also of first order but it tilts on the left of $D_c$, namely the re-orientation transition occurs in a small region below $D_c$, unlike what we find here for the Potts model. To explain the "left tilting" of the Heisenberg case, we have used the following entropy argument:  the Heisenberg in-plane configuration has a spin-wave entropy larger than that of the perpendicular configuration at finite $T$, so the re-orientation occurs in "favor" of the in-plane configuration, it goes from perpendicular to in-plane ordering with increasing $T$. Obviously, this argument for the Heisenberg case  does not apply to the Potts model because we have here the inverse re-orientation transition.  We think that, due to the discrete nature of the Potts spins, spin-waves cannot be excited, so there is no spin-wave entropy as in the Heisenberg case. The perpendicular anisotropy $A$ is thus dominant at finite $T$ for $D$ slightly larger than $D_c$.

\subsection{Thin films}
The case of thin films with a thickness $L_z$ where $L_z$ goes from a few  to a dozen atomic layers has a very similar re-orientation transition as that shown above for the 2D case.

Let us show results for  $J_s=J$ in Figs. \ref{GS3D}-\ref{ECMX} below.  The effect of surface exchange integral $J_s$ will be shown in the following subsection.

Let us show in Fig. \ref{GS3D} the GS obtained by the steepest-descent method with $A=0.5$ and $J=1$ as before, for two thicknesses $L_z=4$ and $L_z=6$.  Changing the film thickness results in changing the dipolar energy at each lattice site. Therefore, the
critical value $D_c$ will change accordingly. We note the periodic layered structures at large $D$ and $r_c$ for both cases.
In the case $L_z=4$, for $r_c=\sqrt{6}$ the critical value $D_c$ above which the GS changes from the perpendicular to the in-plane configuration is $D_c=0.305$.

\begin{figure}
\centering
\includegraphics[width=80mm,angle=0]{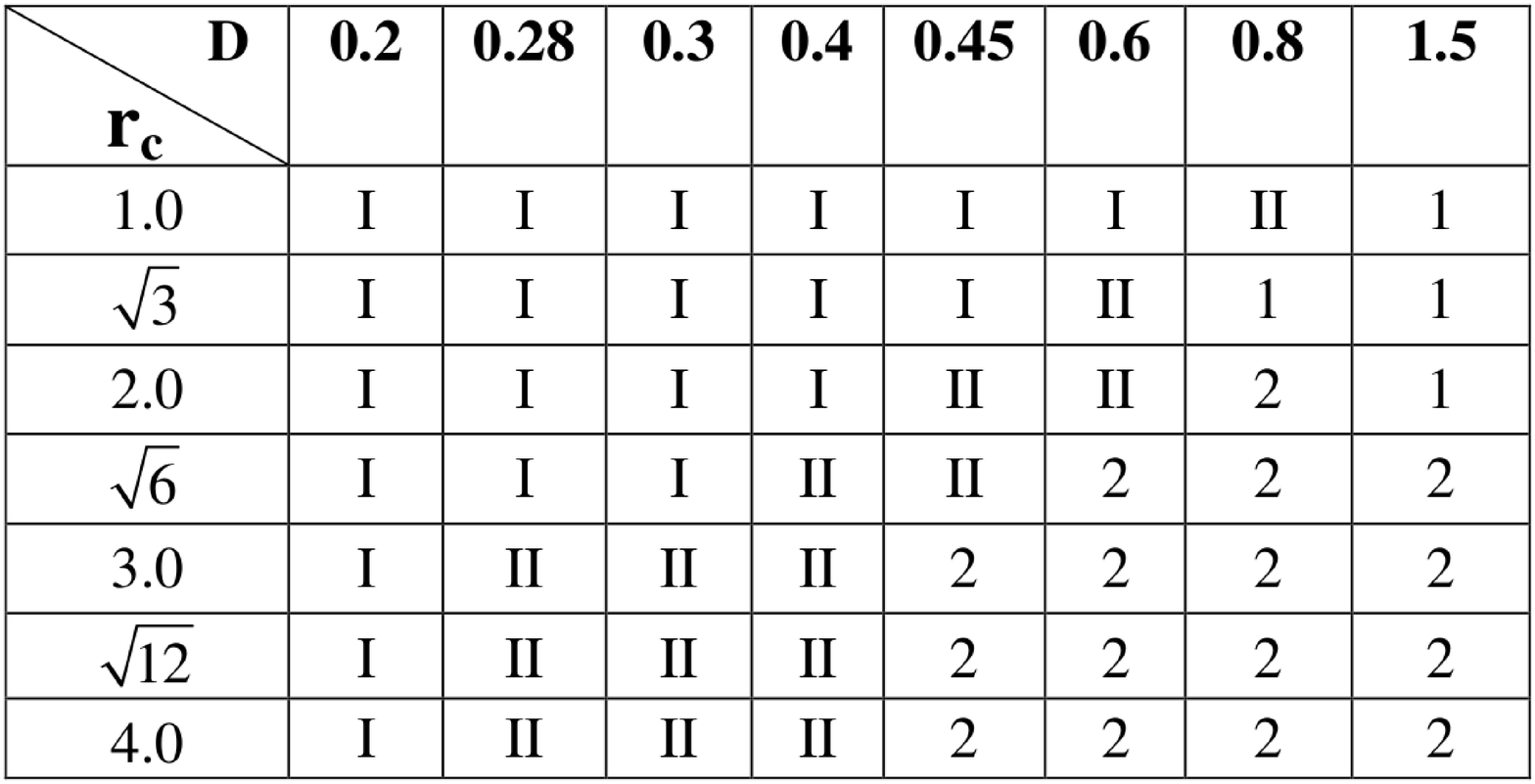}
\includegraphics[width=80mm,angle=0]{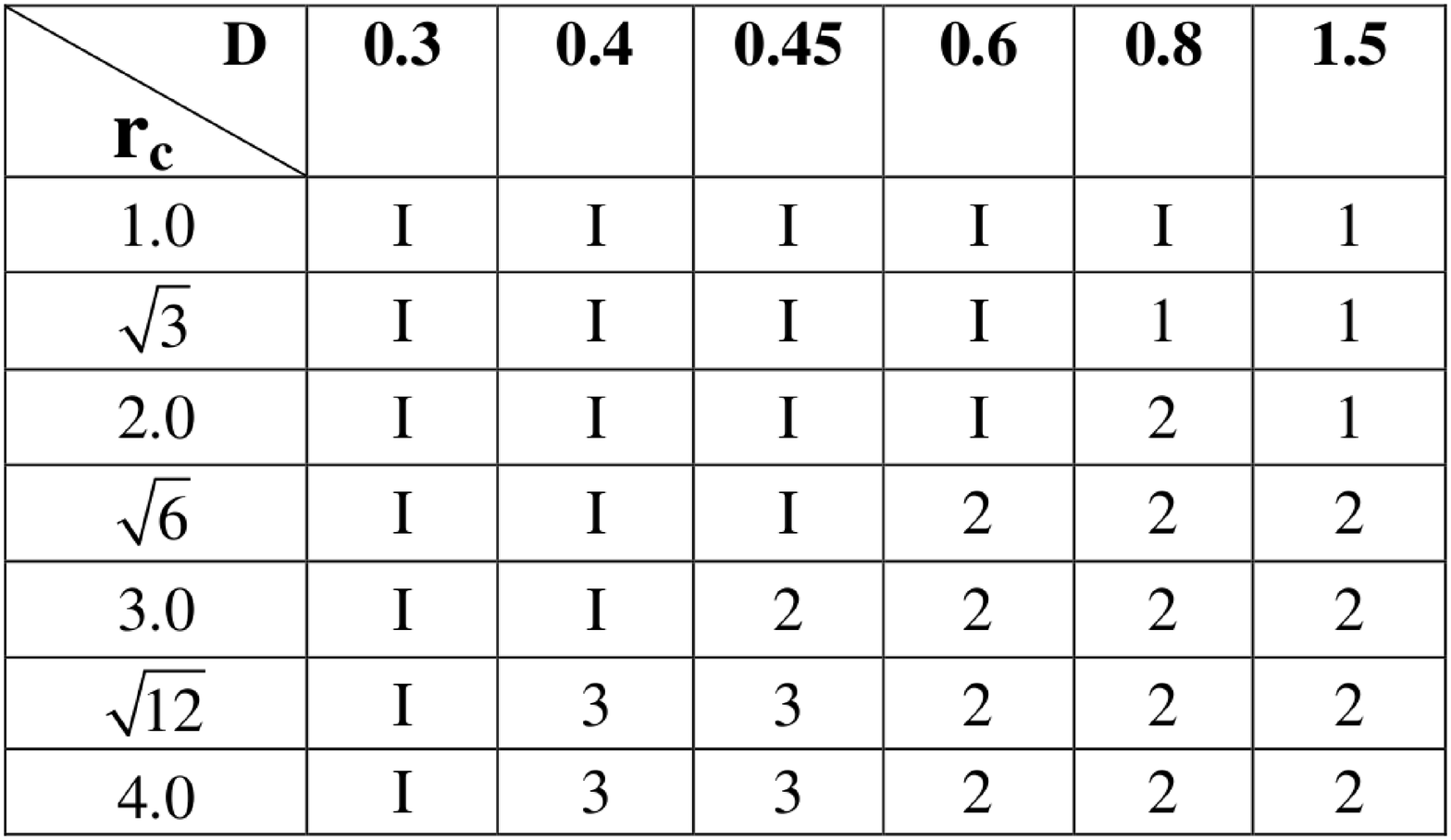}
\caption{Ground states in  a thin film as functions of $(D,r_c)$, for thickness $L_z=4$(top) and 6 (bottom), with $A=0.5$ and $J=1$: the number (I) stands for the perpendicular configuration, the number (II) for the in-plane configuration (spins pointing along $x$ or $y$ axis), the number (1) for alternately one layer in $x$ and one layer in $y$ direction (periodic single-layered structure),  the number (2) stands for the configuration with alternately 2 layers in $x$ alignment and 2 layers in $y$ alignment (periodic bi-layered structure), and the number (3) for alternately three layers in $x$ and three layers in $y$ direction (periodic tri-layered structure) . } \label{GS3D}
\end{figure}

As in the 2D case, we expect interesting behaviors near the critical value $D_c$. For example, when $L_z=4$, $r_c=\sqrt{6}$ and $A=0.5$, we find indeed a re-orientation transition which is shown in Fig. \ref{3DMXMZ}.  The upper curves show clearly a first-order transition from in-plane $x$ ordering to perpendicular ordering at $T\simeq 1.41$.  The total magnetization (middle curve) and the energy (bottom curve) show a discontinuity at that temperature.  The whole phase diagram is shown in Fig. \ref{PD3D}. Note that the line separating the uniform in-plane phase (II) and the periodic single-layered phase (1) is vertical.

\begin{figure}
\centering
\includegraphics[width=50mm,angle=0]{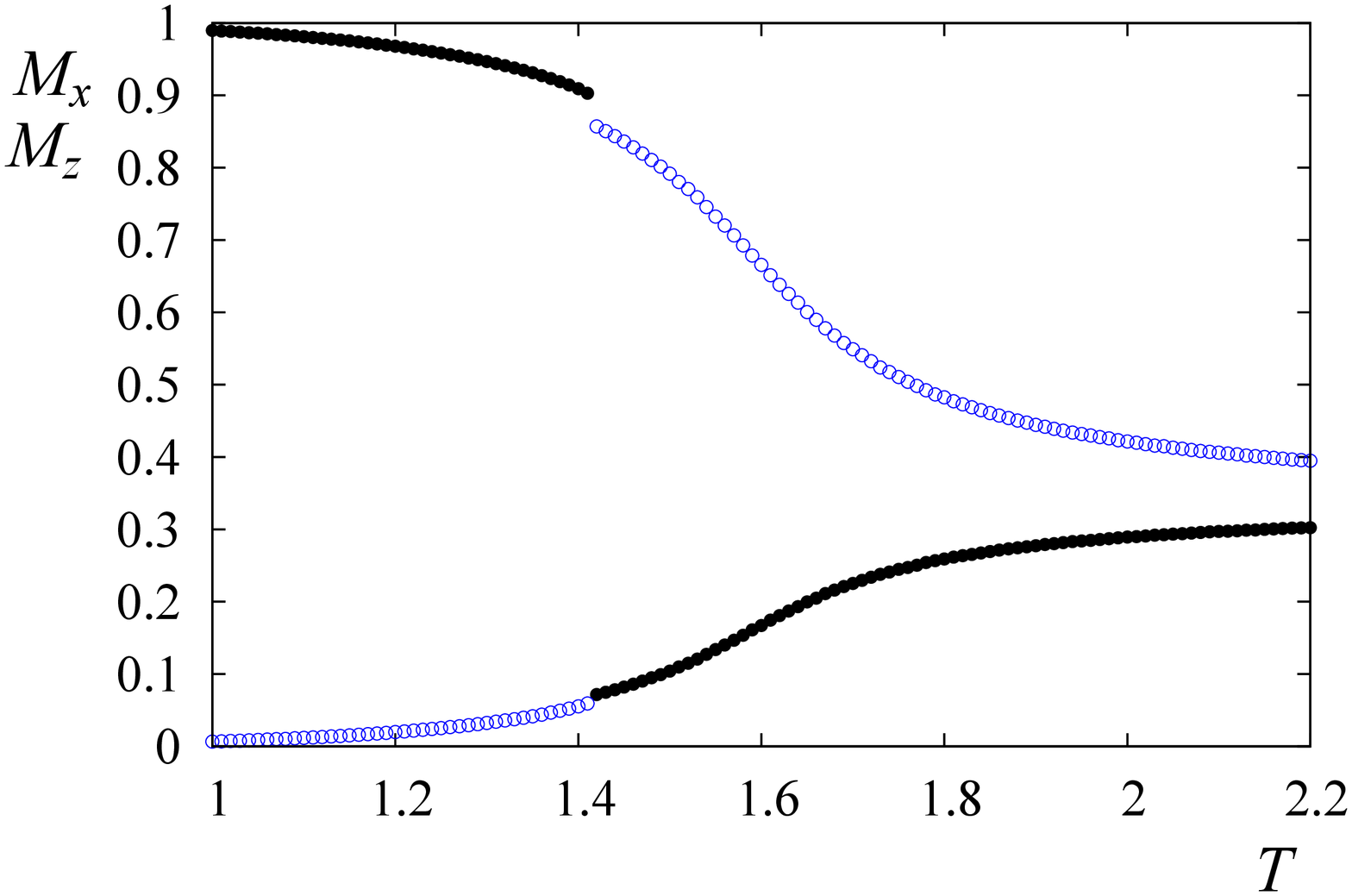}
\includegraphics[width=50mm,angle=0]{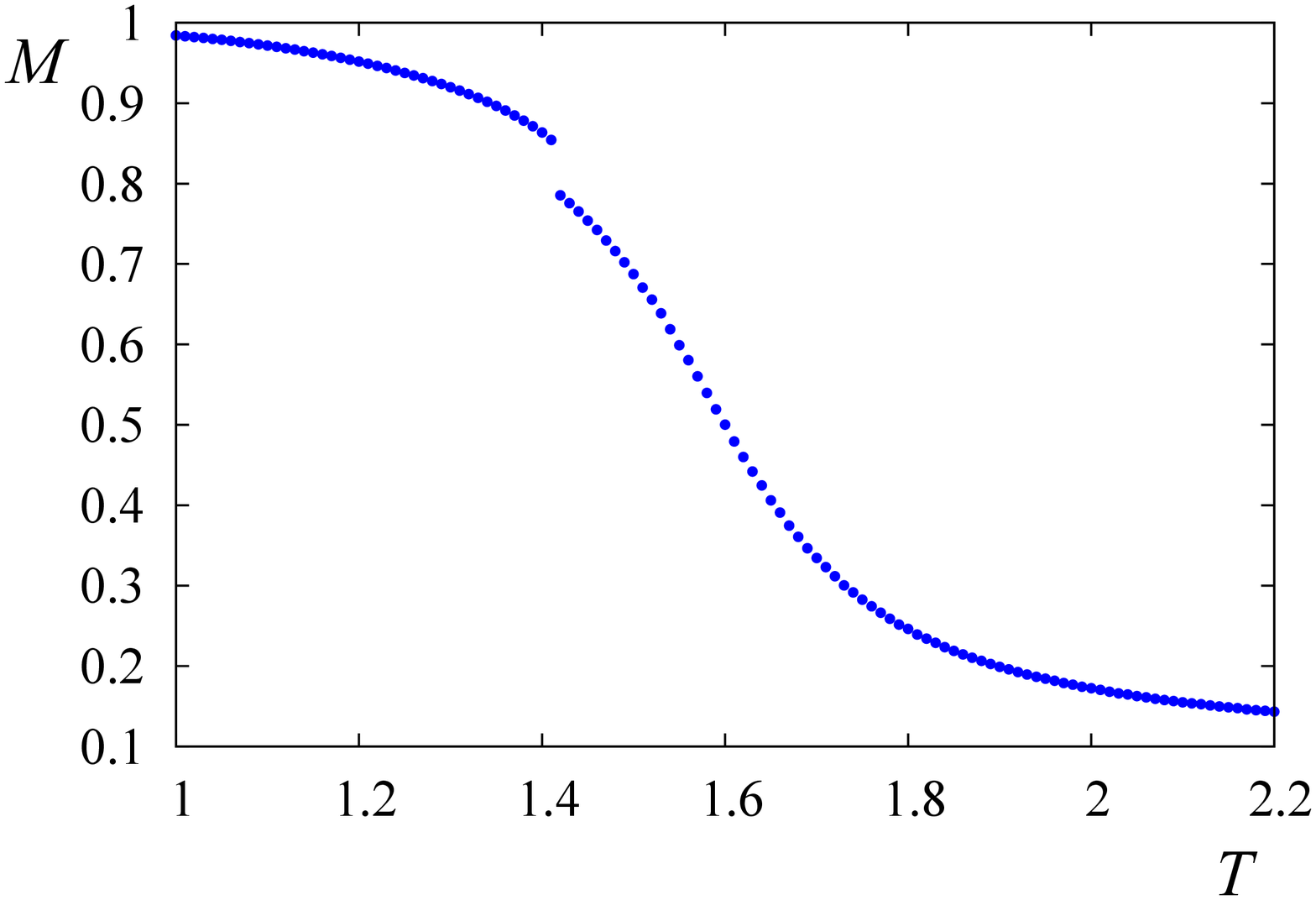}
\includegraphics[width=50mm,angle=0]{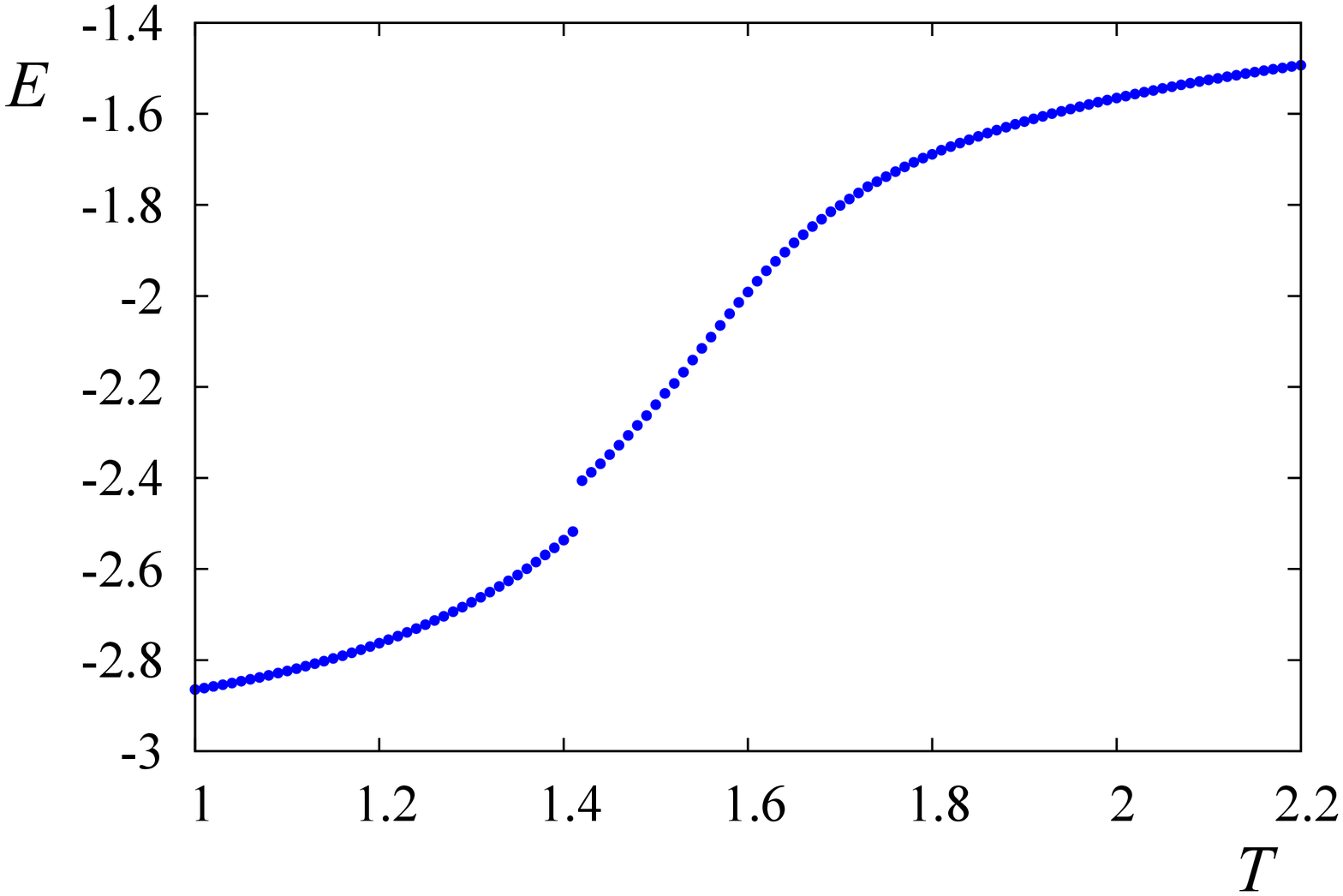}
\caption{(Color online) Film with thickness $L_z=4$ ($L=24$). Top: $M_x$ (black solid circles) and $M_z$ (blue void circles), middle:  total $M$,  bottom: $E$, versus $T$ for $D=0.31$ in the re-orientation transition region. See text for comments.} \label{3DMXMZ}
\end{figure}

\begin{figure}
\centering
\includegraphics[width=70mm,angle=0]{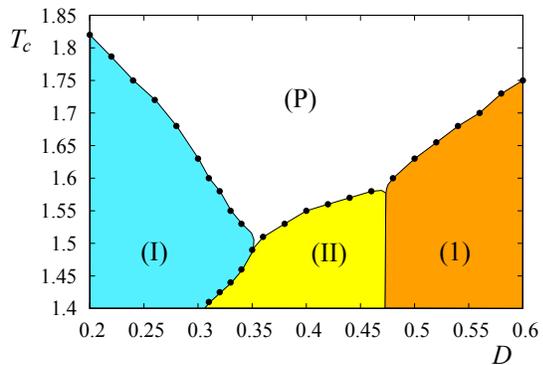}
\caption{(Color online) Phase diagram in thin film of 4-layer thickness: Transition temperature $T_C$ versus $D$, with $A=0.5$, $J=1$ and  $L=24$. Phases (I), (II),(1) and (P) are defined in the caption of Fig. \ref{GS3D}. See text for comments. } \label{PD3D}
\end{figure}

To close this subsection, let us show in Fig. \ref{ECMX} the transition at values of $D$ far from the critical values of $D$. There is only one transition from the ordered phase to the paramagnetic phase. As seen the transition from the in-plane ordering [phases (II) and (1)] to the
 paramagnetic phase is  sharper than that from the perpendicular one [phase (I)],  as in the 2D case.
\begin{figure}
\centering
\includegraphics[width=40mm,angle=0]{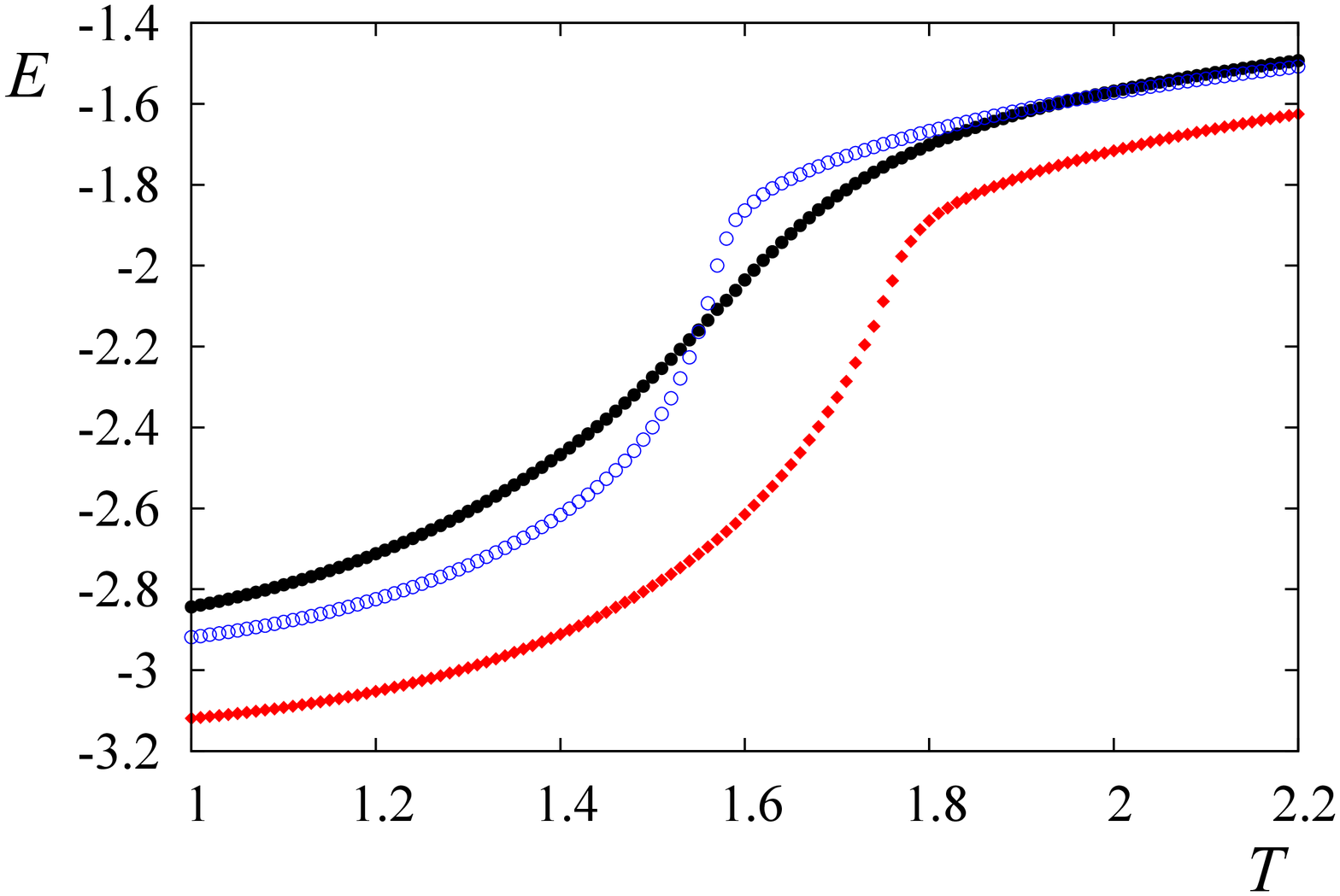}
\includegraphics[width=40mm,angle=0]{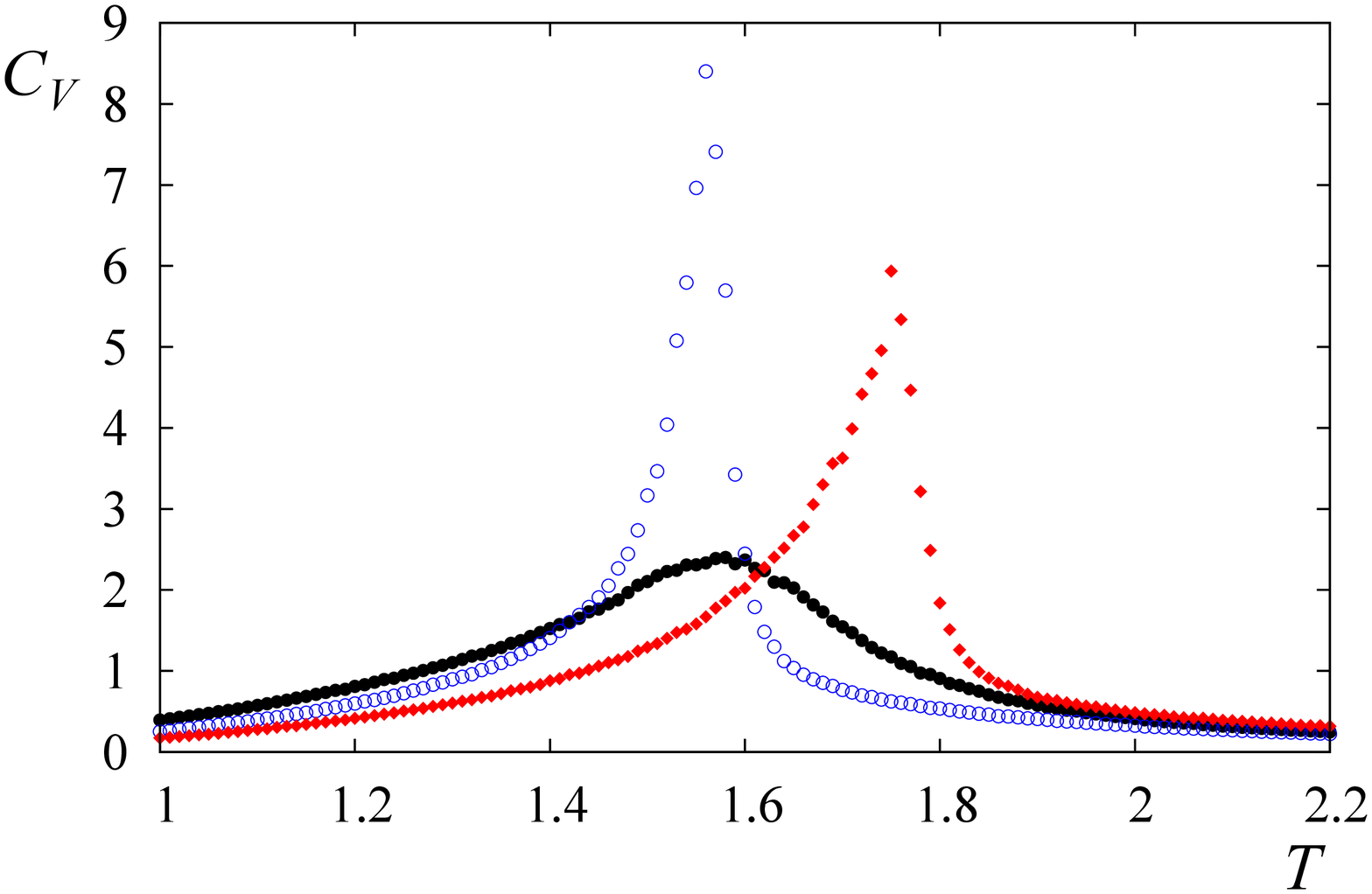}
\includegraphics[width=40mm,angle=0]{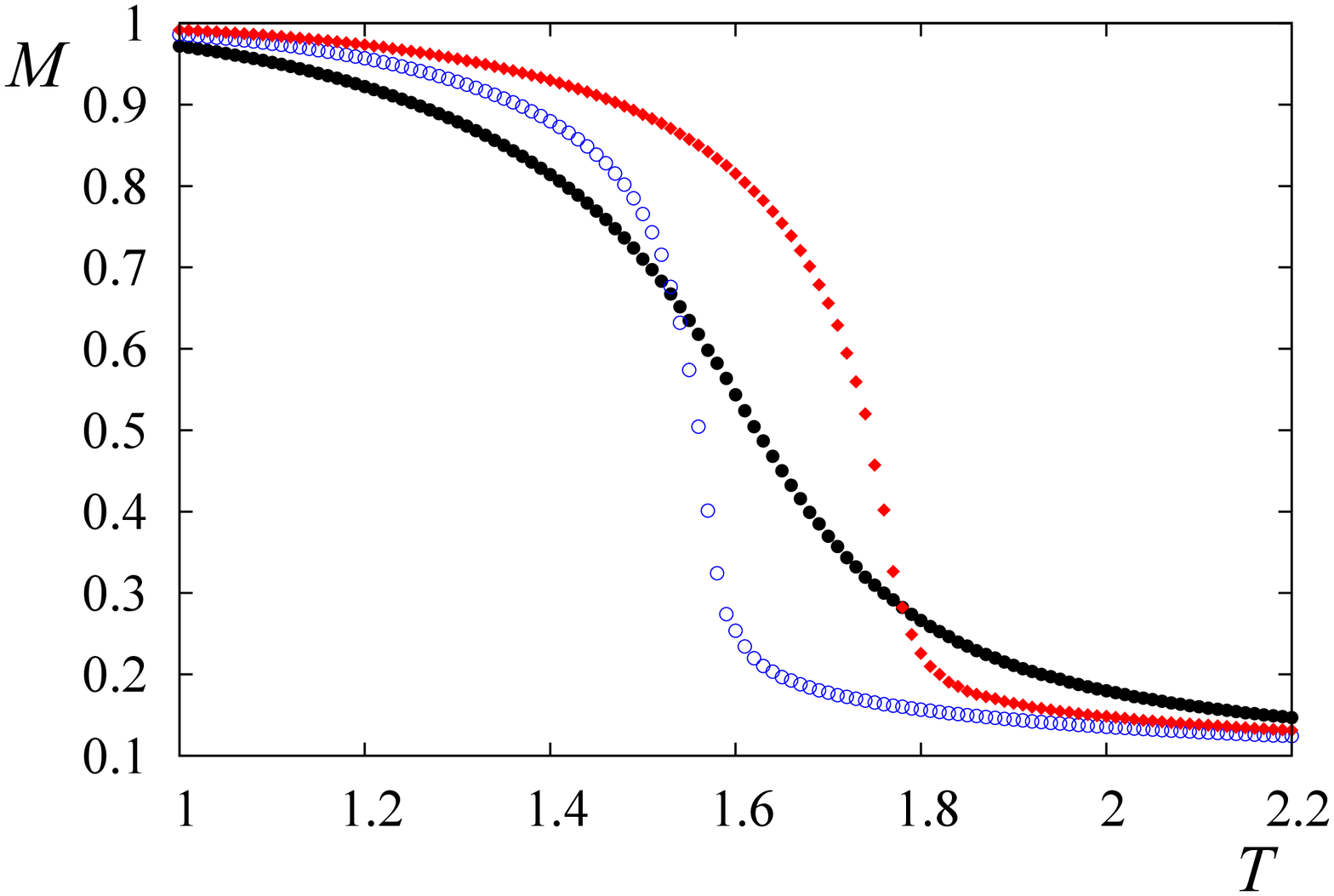}
\includegraphics[width=40mm,angle=0]{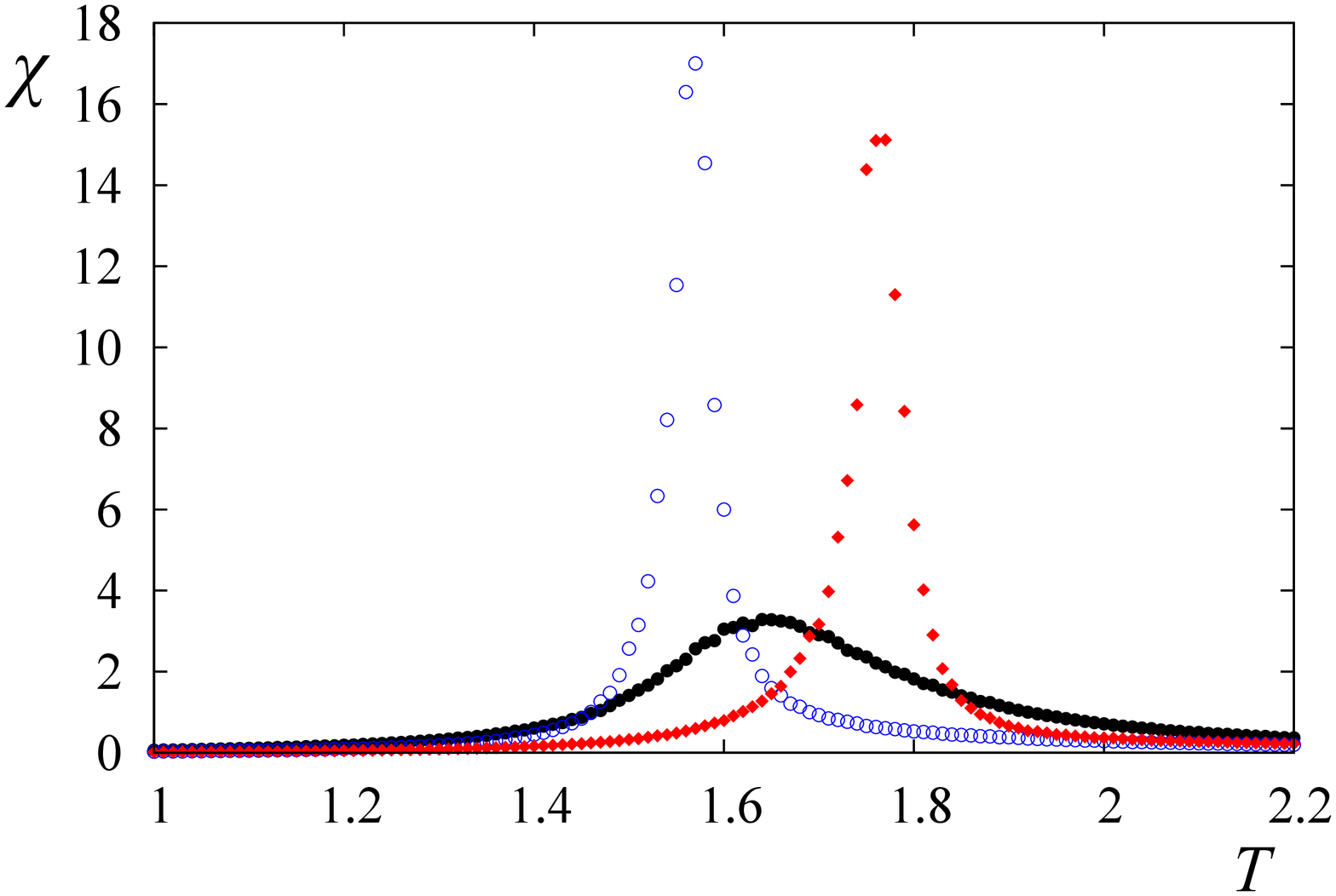}
\caption{(Color online) $E$, $C_V$, $M$ and $\chi$ versus $T$ for $D=$0.3 (black solid circles), 0.4 (blue void circles) and 0.6 (red diamonds), $L_z=4$, $L=24$.} \label{ECMX}
\end{figure}

\subsection{Effect of surface exchange interaction}

\begin{figure}
\centering
\includegraphics[width=40mm,angle=0]{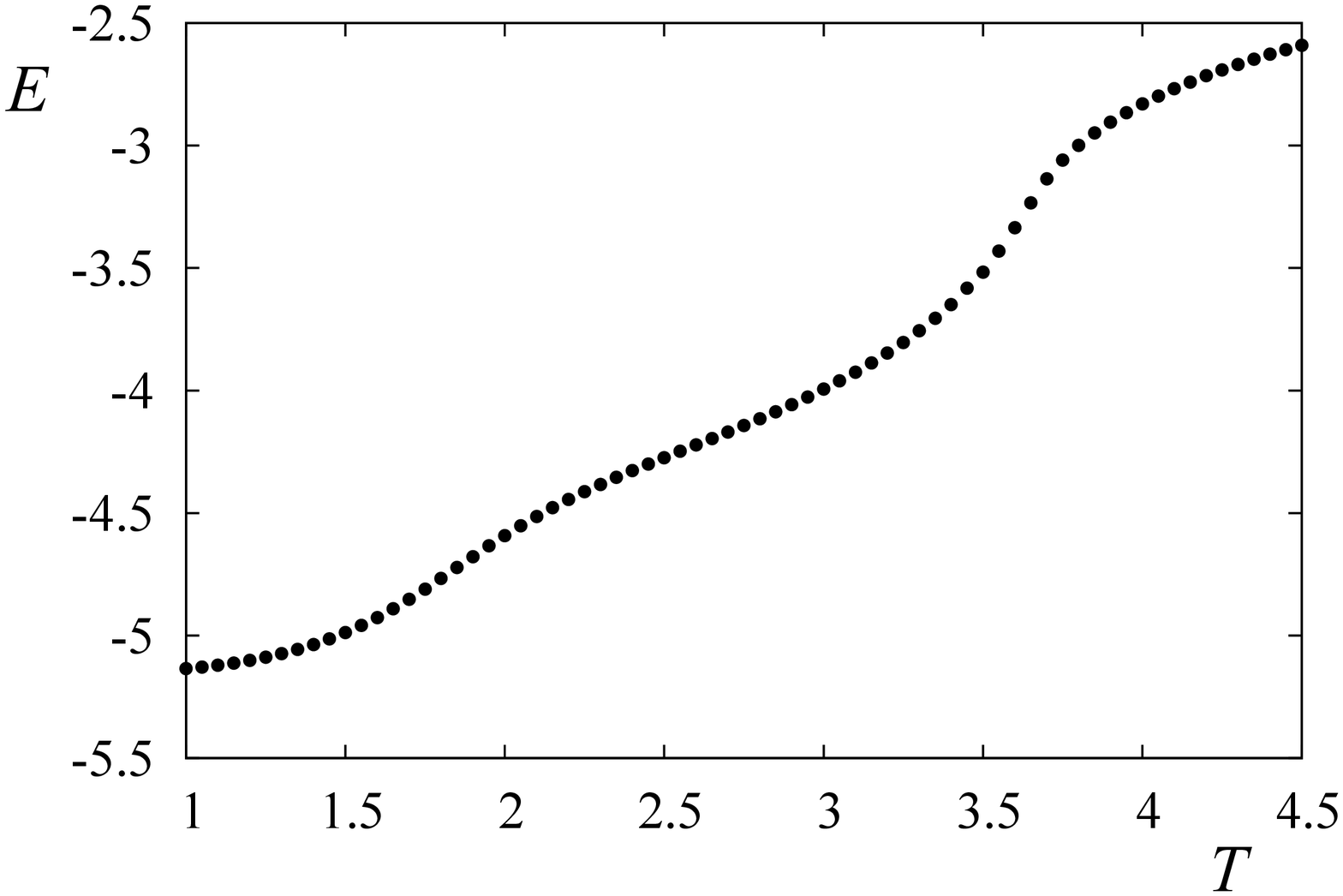}
\includegraphics[width=40mm,angle=0]{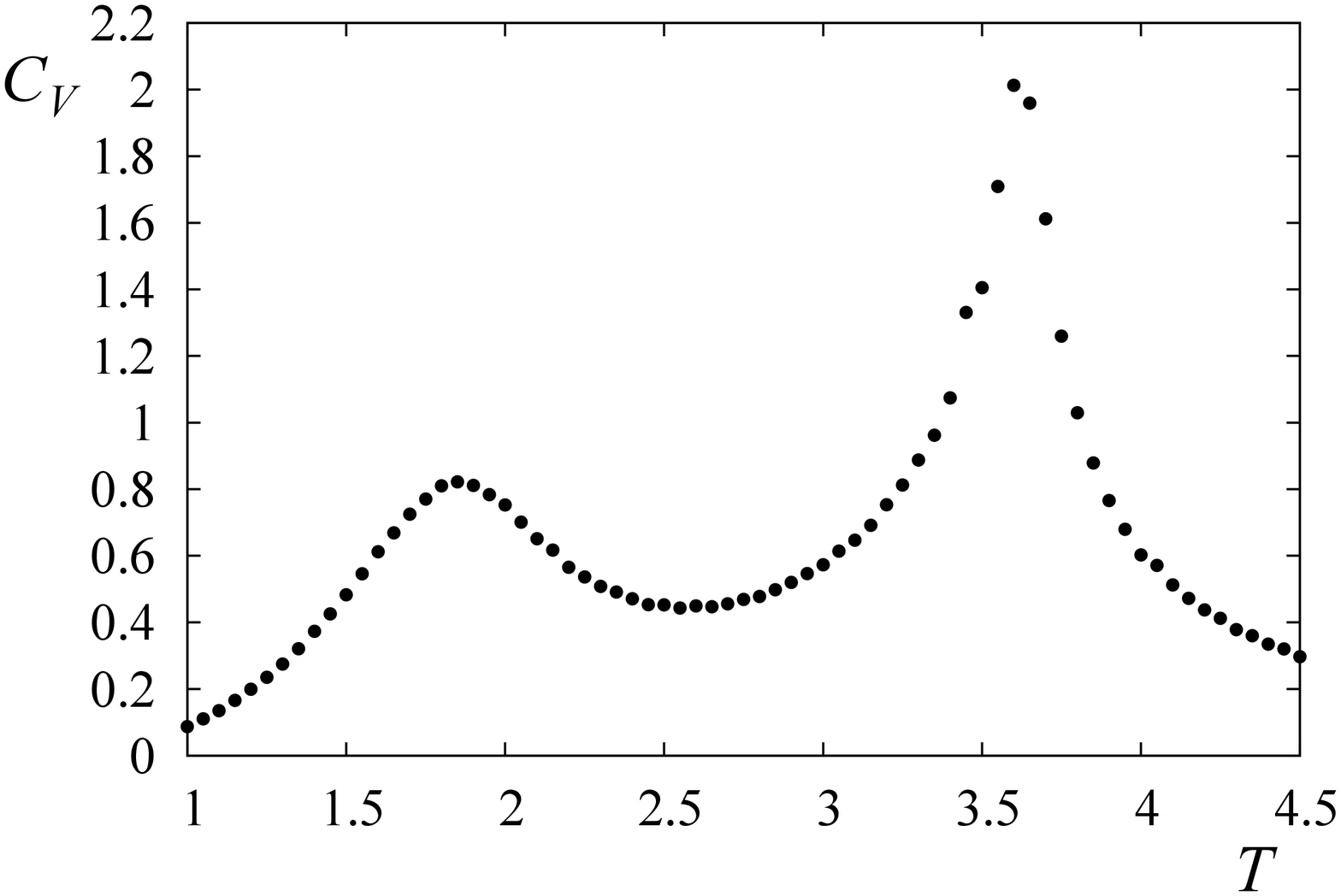}
\includegraphics[width=40mm,angle=0]{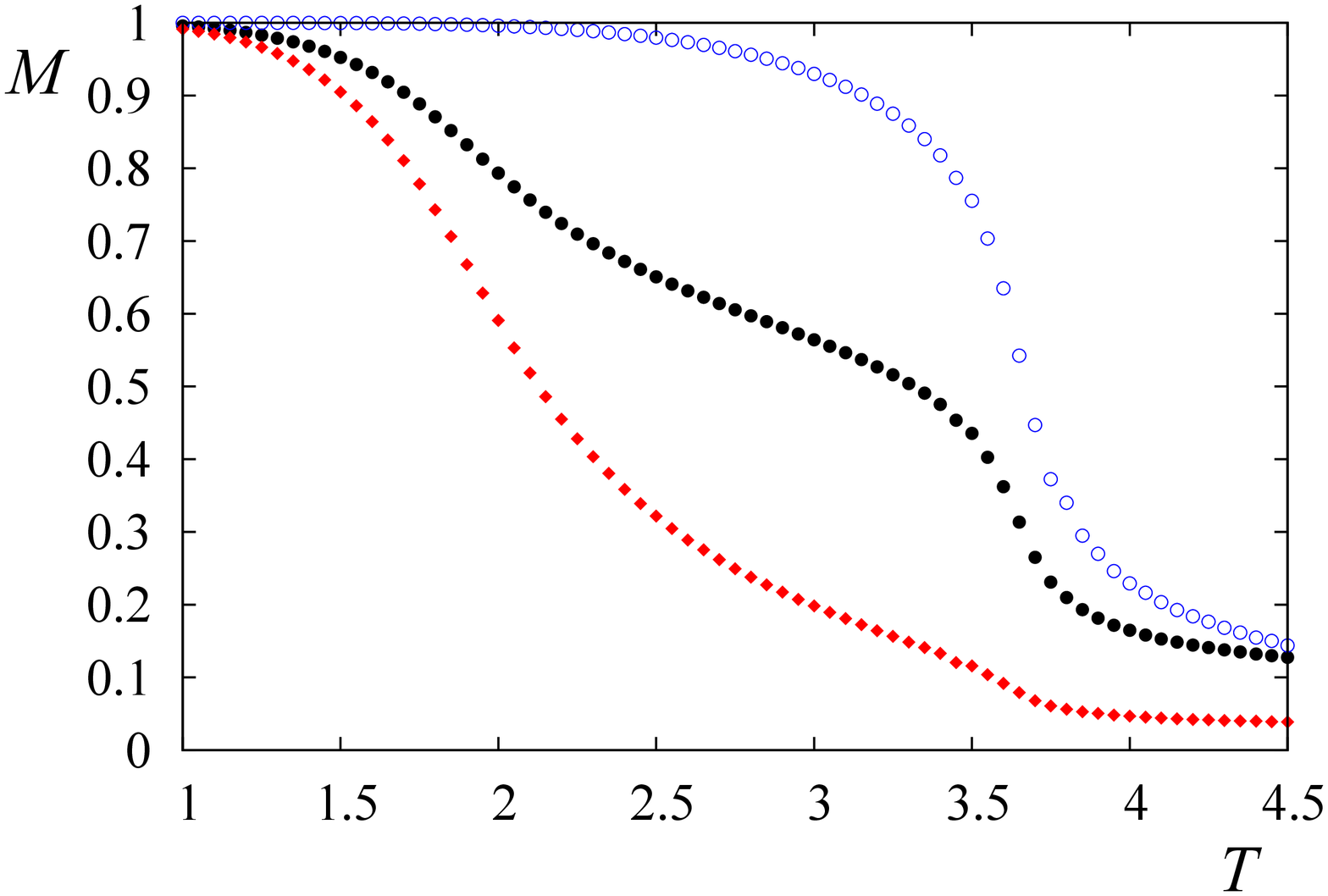}
\includegraphics[width=40mm,angle=0]{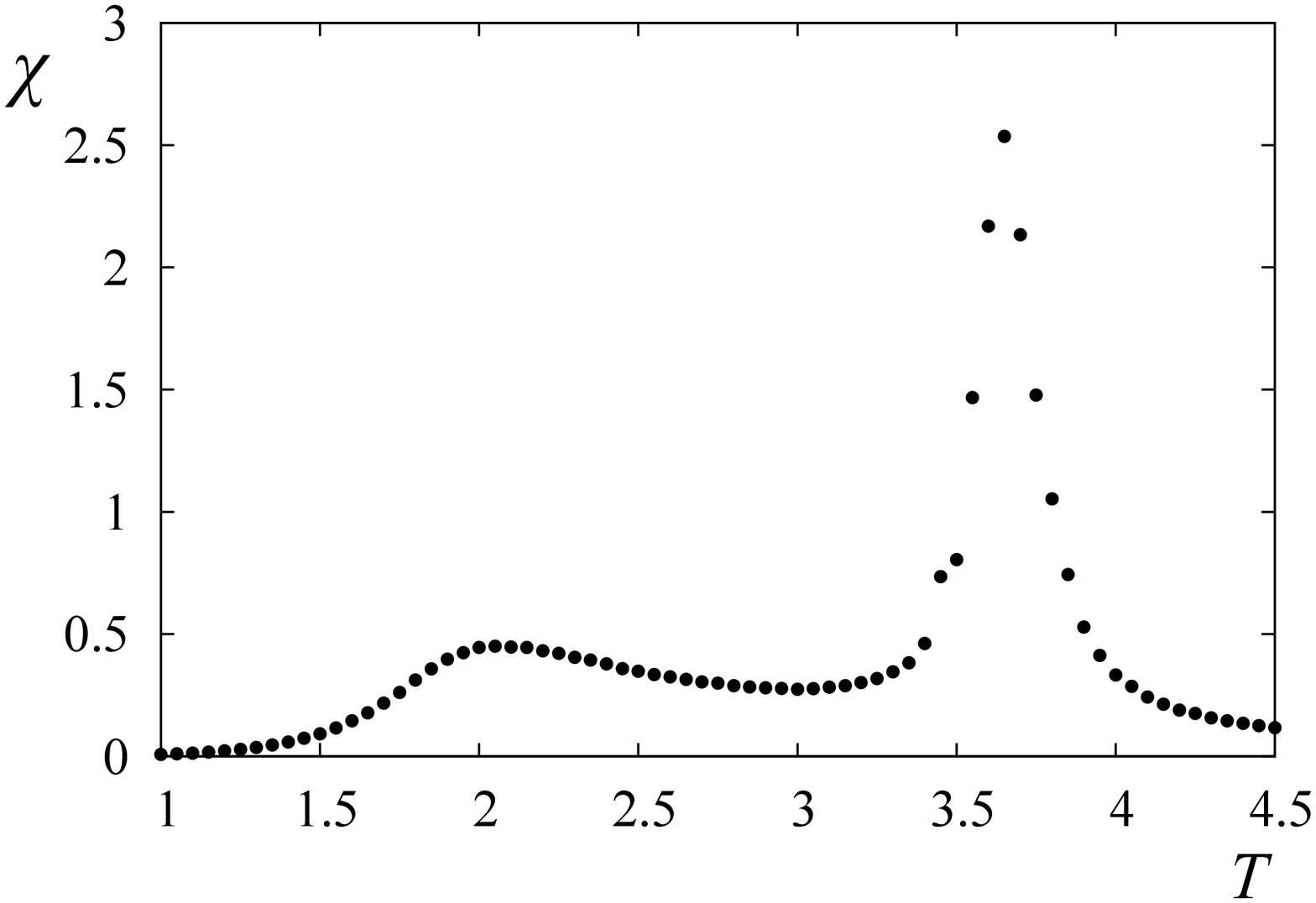}
\caption{(Color online) $E$, $C_V$, $M$ and $\chi$ of a 4-layer film versus $T$ for $D=$0.6 with $J_s=3$ ($L=24$). The surface magnetization is shown by blue void circles, the bulk magnetization by red diamonds and the total curves by black solid circles.} \label{JS}
\end{figure}

We have calculated the effect of $J_s$ by taking its values far from the bulk value ($J=1$) for several values of $D$.  In general, when $J_s$ is smaller than $J$ the surface spins become disordered at a temperature $T$ below the temperature where the interior layers become disordered. This case corresponds to the soft surface (or magnetically "dead" surface layer) \cite{Diep81}. On the other hand, when $J_s>J$, we have the inverse situation: the interior spins become disordered at a temperature lower that of the surface disordering. We have here the case of a magnetically hard surface. We show in Fig. \ref{JS} an example of a hard surface in the case where $J_s=3$ for $D=0.6$ with $L_z=4$. The same feature is observed for $D=0.4$.  Note that the surface and bulk transitions are seen by the respective peaks in the specific heat and the susceptibility.  In the re-orientation region, the situation is very complicated as expected because the surface transition occurs in the re-orientation zone.

\subsection{Discussion}
Note that for a given $D$, the effect of the cutoff distance $r_c$ is to move the critical value of $D_c$ as seen in Figs. \ref{GS2D} and \ref{GS3D}.   At $r_c=\sqrt{10}\simeq 3.16$ one has 146 neighbors for each interior spin (not near the surface).  This huge number makes MC simulations CPU-time consuming.  We therefore performed simulations at finite $T$ only with two values of $r_c$ in the 2D case.  As seen in Fig. \ref{PD2D}, the change of $r_c$ does not alter our conclusion on the re-orientation transition.
 We think that the cutoff  is more than a technical necessity, it involves also physical reality. We have in mind the observation that in most experimental systems  interaction between faraway neighbors can be neglected. The concept that the interaction range between particles can go to infinity is a theoretical concept. Models in statistical physics limited to interaction between NN are known to interpret with success experiments \cite{Zangwill,Zinn}. Rarely we have to go farther than third NN. For example, in our recent paper on the spin resistivity in semiconducting MnTe, we took interactions up to third neighbors to get an excellent agreement with experiments \cite{MagninMnTe}.  Therefore, we wanted to test  in the present paper how physical results depend on $r_c$ in the dipolar interaction. If we know for sure that in a thin film the interaction is dipolar and that a double-layered structure for example is observed, from what is found above we can suggest the interaction range between spins in the system.
Finally, we note that if we change $A$, the value of $D_c$ will change.  The choice of $A$=0.5 which is a half of $J$ is a reasonable choice  to make the re-orientation happen.  A smaller $A$ will induce a smaller $D_c$ but again, the physics found above will not change.

 \section{Concluding Remarks}\label{conclu}

 We have shown in this paper MC results on the phase transition in thin magnetic films using the Potts model including a short-range exchange interaction $J$ and a  long-range dipolar interaction of strength $D$, truncated at a distance $r_c$.  We have also included a perpendicular anisotropy $A$ which is known to exist in very thin films.

 Among the striking results, let us mention the re-orientation transition which occurs in 2D and in thin films at a finite temperature below the overall disordering. This re-orientation is a very strong first-order transition as seen by the discontinuity of the energy and the magnetization.  We emphasize that the re-orientation is possible only because we have two competing interactions: the perpendicular anisotropy and the dipolar interaction.

 We would like to acknowledge the financial support from a grant of the binational cooperation program POLONIUM of the French and Polish Governments.

{}

\end{document}